\documentclass[12pt,preprint]{aastex}
\usepackage[dvips]{color}
\usepackage{longtable,lscape}
\usepackage{amsmath}
\usepackage[usenames,dvipsnames]{xcolor}

\begin{document}
\newcommand{\wcen}  {$\omega$~Cen}
\newcommand{\boo}   {Bo\"{o}tes~I}
\newcommand{\boos}  {Boo-1137}
\newcommand{\kms}   {\rm km~s$^{-1}$} 
\newcommand{\teff}  {$T_{\rm eff}$} 
\newcommand{\logg}  {$\log g$} 
\newcommand{\loggf} {log~$gf$} 
\newcommand{\bvz}   {$(B-V)_{0}$} 
\newcommand{\feh}   {[Fe/H]}
\newcommand{\alphafe} {[$\alpha$/Fe]}
\newcommand{\vxi}   {$\xi_t$}
\newcommand{\uf}    {ultra-faint}
\newcommand{\mv}    {$M_{V}$}
\newcommand{\ebv}    {$E(B-V)$}
\newcommand{\ebi}    {$E(B-I)$}
\newcommand{\bi}    {$B-I$}
\newcommand{\bio}   {$(B-I)_{0}$}
\newcommand{\vhb}   {$V_{\rm HB}$}
\newcommand{\msun} {\rm M$_\odot$}
\newcommand{\ltsima} {$\; \buildrel < \over \sim \;$}
\newcommand{\simlt} {\lower.5ex\hbox{\ltsima}}
\newcommand{\gtsima} {$\; \buildrel > \over \sim \;$}
\newcommand{\simgt} {\lower.5ex\hbox{\gtsima}}

\title{THE POPULATIONS OF CARINA. I.  DECODING THE COLOR--MAGNITUDE DIAGRAM}

\author{JOHN E. NORRIS\altaffilmark{1}, DAVID YONG\altaffilmark{1},
  KIM A. VENN\altaffilmark{2}, AARON DOTTER\altaffilmark{1}, LUCA
  CASAGRANDE\altaffilmark{1}, AND GERARD GILMORE\altaffilmark{3} }

\altaffiltext{1}{Research School of Astronomy and Astrophysics, The
  Australian National University, Canberra, ACT 2611, Australia;
  jen@mso.anu.edu.au, yong@mso.anu.edu.au, aaron.dotter@gmail.com,
  luca@mso.anu.edu.au}

\altaffiltext{2}{Department of Physics and Astronomy, University of 
Victoria, 3800 Finnerty Road, Victoria, BC V8P 1A1, Canada; kvenn@uvic.ca}

\altaffiltext{3}{Institute of Astronomy, University of Cambridge,
  Madingley Road, Cambridge CB3 0HA, UK; gil@ast.cam.ac.uk}

\begin{abstract}

We investigate the color-magnitude diagram (CMD) of the Carina dwarf
spheroidal galaxy using data of \citet{stetson11} and synthetic CMDs
based on isochrones of \citet{dotter08}, in terms of the parameters
[Fe/H], age, and {\alphafe}, for the cases when (i) {\alphafe} is held
constant and (ii) {\alphafe} is varied.  The data are well described
by four basic epochs of star formation, having [Fe/H] = --1.85, --1.5,
--1.2, and $\sim$--1.15 and ages $\sim$~13, 7, $\sim3.5$, and
$\sim$1.5 Gyr, respectively (for {\alphafe} = 0.1 (constant
{\alphafe}) and {\alphafe} = 0.2, 0.1, --0.2, --0.2 (variable
{\alphafe})), with small spreads in [Fe/H] and age of order 0.1 dex
and 1 -- 3 Gyr.  Within an elliptical radius 13.1{$\arcmin$}, the mass
fractions of the populations, at their times of formation, were (in
decreasing age order) 0.34, 0.39, 0.23, and 0.04.  This formalism
reproduces five observed CMD features (two distinct subgiant branches
of old and intermediate-age populations, two younger, main-sequence
components, and the small color dispersion on the red giant branch
(RGB). The parameters of the youngest population are less certain than
those of the others, and given it is less centrally concentrated it
may not be directly related to them. High-resolution spectroscopically
analyzed RGB samples appear statistically incomplete compared with
those selected using radial velocity, which contain bluer stars
comprising $\sim$5 -- 10\% of the samples.  We conjecture these
objects may, at least in part, be members of the youngest population.
We use the CMD simulations to obtain insight into the population
structure of Carina's upper RGB.

\end{abstract}


\section{INTRODUCTION} \label{sec:intro}

Since its discovery four decades ago by \citet{cannon77}, the Carina
dwarf spheroidal (dSph) galaxy has been the subject of enormous
interest -- driven by what it has to tell us about Carina's formation
and chemical enrichment, and the evolution of structure in the early
Universe.  Table~\ref{tab:highlights} presents a list of some 20 major
milestones that have contributed to our current understanding of the
system from the above viewpoints\footnote{Where possible, in the
  second column of Table~\ref{tab:highlights} the first number refers
  to the discovery/seminal paper, while subsequent references refer to
  further important contributions.}.  For recent comprehensive
descriptions of the development of the extensive literature on Carina,
we refer the reader to \citet{deboer14} and \citet{kordopatis16}, and
for an introduction to matters related to Carina's population and
abundance structures see \citet{tolstoy09} and \citet{venn12}.

We begin here by briefly summarizing aspects of the available material
that inform the impetus for the present work.  In particular there is
clear agreement that the majority of Carina's stars are of
intermediate age, with a smaller and well-defined older component.  To
cite \citet{deboer14} ``Two main episodes of star formation occurred
at old ($>$8~Gyr) and intermediate (2 -- 8 Gyr) ages, both enriching
stars starting from low metallicities ([Fe/H] $<$ --2~dex)''.  That
said, and as the reader will see in Table~\ref{tab:highlights}, three
and as many of six populations have also been suggested.  Our starting
point is that the extremely high quality photometric color magnitude
diagram (CMD) of \citet{bono10} and \citet{stetson11} suggest to us
that there were essentially four basic episodes of star formation,
which we shall describe in Section~\ref{sec:decoded}.  The excellence
of the photometry and the observed tightness of Carina's RGB lead to
very strong constraints on age and chemical abundance distributions.
Second, while the structure of the CMD is fundamentally driven by
these two parameters, the CMD alone currently has the power to
accurately constrain the age distribution function of the system.  On
the other hand, the chemical abundances ([Fe/H] and [X/Fe]) needed to
best determine the complete description of the system are provided by
the brighter stars on the RGB.  In this case, however, in contrast to
the excellence of the photometric data, the available high-resolution
spectroscopy is somewhat compromised by the faintness of Carina's RGB
and currently of insufficient signal-to-noise ($S/N$) and too sparse
and inhomogeneously analyzed to provide a definitive discussion of its
chemical development and evolutionary history.  To more fully address
this issue we have sought to augment and homogeneously analyze
high-resolution spectroscopic data to produce chemical abundances for
a large sample of stars on Carina's RGB.  To this end we have
assembled homogeneously determined relative abundances for 19 elements
in 63 of its red giants.

We separate our investigation into two papers.  The first, presented
here, contains an analysis of the CMD.  In the second
(\citealp[hereafter referred to as Paper II]{norris17b}), we shall
present the abundance results for the 63 red giants, which will be
discussed in the light of age and abundance constraints from our
analysis of the CMD in order to interpret the details of the above
putative four populations that are present in the upper RGB sample.

\subsection{Outline of the Present Work}

In Section~\ref{sec:decoded} we present the very high quality $BVI$
photometry of Carina described by \citet{stetson11} and made available
by P.B. Stetson (2014, private communication), and discuss the
morphology of its ($V$, \bi) CMD in terms of the distinctive features
that we shall use to constrain what we consider to be its four basic
populations and their epochs of formation.  We also compare this CMD
with equally high quality data for the globular clusters M13, M92, and
{\wcen}.  A comparison with the isochrones of \citet{dotter08}
confirms the large differences between the intracluster age and
abundance distributions of the globular clusters, on the one hand, and
Carina, on the other.  We discuss the implications of these differences
for [Fe/H] values determined for Carina from observations of the Ca II
infrared triplet and calibrations based on Galactic globular clusters.
Section~\ref{sec:cmd} contains the core of our investigation, in which
we present synthetic CMDs that reproduce the distinctive features of
Carina's CMD.  These lead to estimates of the relative mass fractions
of the four populations.  Section~\ref{sec:upperrgb} presents a
discussion of the synthetic distribution functions of metallicity
([Fe/H]), {\alphafe}, and age for stars on Carina's upper RGB, which
permit insight into the relative contributions of the four
populations.  We shall compare these results in Paper II with the
observations of the sample of 63 Carina red giants based on our
high-resolution spectroscopic abundance analyses.  The
results are summarized in Section~\ref{sec:discussion}.

\section{PHOTOMETRY OF CARINA AND THE GALACTIC GLOBULAR CLUSTERS}\label{sec:decoded}

\subsection{The Carina CMD}\label{sec:carina_cmd}

Figure~\ref{fig:carcmd} presents the data of Stetson (2014, private
communication) in the ({\mv}, {\bio}) -- plane for all of the $\sim
31000$ stars within an elliptical region centered on RA = 06 41 36 and
Dec = --50 58 26 (2000) and elliptic radius r$_{\rm ellip}$ =
13.1{$\arcmin$,} with e = 0.33 and position angle 65$^{\circ}$
\citep{irwin95}.  To determine {\mv} and {\bio} we adopted {\ebv} =
0.06 and (m -- M)$_V$ = 20.05, following \citet{venn12} and adopting
{\ebi} = 2.4$\times${\ebv}.  The left panel is the CMD, while in the
right panel we show the Hess diagram, using contours.  In what
follows, we focus on the following five basic morphological aspects of
this CMD: (i) the unexpected tightness of the RGB ($\sigma$({\bi})
$\sim 0.07$ over the range --2.5 $<$ {\mv} $<$ --1.0) (see
Section~\ref{sec:tightness}) given the large observed dispersion in
iron abundance ($\sigma$[Fe/H] $\sim 0.4$ in our analysis for stars in
this magnitude interval); (ii) the lower subgiant branch (SGB), at
{\mv} $\sim~3.3$; (iii) the upper SGB, at {\mv} $\sim~2.7$; (iv) the
main sequence stub above the upper SGB, at ({\mv}, {\bio}) = (2.3,
0.5); and (v) the upper, bluer, and somewhat ephemeral (presumably)
main sequence at ({\mv}, {\bio}) = (2.0, 0.0) -- (3.0,
0.2)\footnote{One might wonder whether Carina's horizontal branch (HB)
  reaches blueward and fainter into the region of Carina's upper main
  sequence and subgiant branch, and might contribute to what we have
  designated the ``fourth population''.  We are of the view that it is
  unlikely to play a major role, for two reasons.  First, in Galactic
  globular clusters with the most blueward and fainter extended HBs,
  such as {\wcen} and M13 (see the results of \citet{bellini09} and
  Stetson
  (http://www.cadc-ccda.hia-iha.nrc-cnrc.gc.ca/en/community/STETSON/standards/)
  in our Figure~\ref{fig:cargc}) and NGC~6752 (see
  \citealp{campbell13} based on data of \citealp{grundahl99}), the HBs
  reach down to {\mv} $\sim$ 2.0 -- 2.5, in contrast to Carina's
  fourth population, which is clearly seen at {\mv} = 3.0.  Second,
  inspection of our Figure~\ref{fig:cargc} shows a very different
  {\mv} density distribution of the blue horizontal branch (BHB) stars
  of $\omega$~Cen and M13 as one moves from bright to fainter
  magnitudes than that seen for the collective BHB and fourth
  populations of Carina.  An alternative suggestion is that the fourth
  population is made up of blue stragglers \citep{santana16}. We shall
  return to these points in Section~\ref{sec:spreads}.}.  We shall
assume that each of features (ii) -- (v) corresponds to a distinct
stellar population and epoch of star formation in Carina, and that the
parameters of these populations together produce a very tight RGB
(feature (i)).  In what follows we refer to stars in (ii) -- (v)
chronologically -- as the ``first'', ``second'', ``third'', and
``fourth'' populations, respectively.

\subsubsection{The Tightness of Carina's Upper RGB}\label{sec:tightness}

In order to discuss the tightness of the upper RGB, we plot the
Stetson et al. photometry of Carina's upper CMD in
Figure~\ref{fig:uppercmd}, highlighting the radial-velocity members
from three independent investigations.  In all three panels the small
points represent all stars observed.  These have been overplotted in
the left panel with large red symbols for the 63 red giants having
chemical abundances based on high-resolution model-atmosphere analysis
(Paper II).  In the middle panel the large green symbols stand for the
radial-velocity members (at the 2.5$\sigma$ level) from
\citet{koch06}; and on the right the large yellow symbols represent
the radial-velocity members of \citet{walker09}.  In the three
samples, for stars brighter than $M_{V}$ = --1.0, the RMS dispersions
in {\bio} about the quadratic least-squares fits (the black lines in
the three panels) are 0.072 $\pm$ 0.007~mag for the Paper II sample
(59 objects), 0.117 $\pm$ 0.009~mag for the \citet{koch06} sample (83
objects), and 0.100 $\pm$ 0.007~mag for the \citet{walker09} sample
(103 objects), respectively.

An intriguing conclusion from these data is that the RGB width of the
Paper II sample is narrower than those of \citet{koch06} and
\citet{walker09} by some 4$\sigma$ and 3$\sigma$, respectively.  A
likely cause of these differences is that while the results of Koch et
al. and Walker et al. are based on statistically complete samples,
that of Paper II comprises objects chosen to lie on the RGB in
Carina's CMD from a number of works, and is not complete.  Roughly
speaking, in the above magnitude range, the Paper II sample has a
deficit of members blueward of Carina's well-defined RGB at the
$\sim$5 -- 10\% level.  The most obvious explanation of these bluer
stars is that they are asymptotic giant branch (AGB) stars.  In
comparison, for globular cluster giant branch stars with {\mv}
{\simlt}~--1.0, the AGB comprises some 20 -- 30\% of the upper giant
branch (e.g., for M13 and M92 see our Figure~\ref{fig:cargc}; for M5,
\citealp{sandquist04}; and for NGC~6752, \citealp{campbell13}).
Furthermore, the existence of some nine carbon stars in Carina
\citep{azzopardi86} is consistent also with the presence of
intermediate-age AGB stars.  We shall return in
Section~\ref{sec:approx} to the contaminating role of AGB stars in our
comparison between the observed and synthetic CMDs of Carina, and in
Section~\ref{sec:raddis} we shall conjecture, based on our synthetic
CMDs, that in some part they may be relatively-young (age $\sim
2$~Gyr) and on their first ascent of the RGB. Another explanation,
suggested by \citet{lemasle12}, in the context of spectroscopically
selected samples observed at less that optimum $S/N$, is that there is
an abundance bias in that stars of lower abundance are less likely to
be analyzed than those having higher abundance.

Insight into the significance of the above color spreads is provided
by the stellar isochrones of \citet{dotter08}, which tell us that at
{\mv} = --1.8 on the RGB of the isochrone [Fe/H]/{\alphafe}/Age =
--1.5/+0.1/7.0 (very roughly applicable to Carina; see
\citealp{deboer14} and our Figure~\ref{fig:carsyn1}) that the
dispersion $\sigma${\bio} = 0.072 (where we have ignored the very
small observational photometric errors at this magnitude) could be
caused by a dispersion in [Fe/H] of $\sim$0.12 dex.  This is well
below the observed dispersion of $\sigma$[Fe/H] = 0.33 in the
high-resolution spectroscopic abundances of the 63 star sample of
Paper II.  Confirmation of this effect is provided at fainter
magnitudes on the RGB.  In the red box in Figure~\ref{fig:uppercmd},
which lies below the above samples and spans the range 0.0 $<$ {\mv}
$<$ 1.0, the observed RMS dispersion in {\bio} about the quadratic
best fit to the RGB in {\bio} is 0.038 mag (from 405 stars) (after a
small quadratic correction of 0.014 mag to allow for random
observational errors).  For the above isochrone, this could be caused
by a dispersion in [Fe/H] of $\sim 0.06$~dex.  As with the brighter
sample considered above, this value is significantly smaller than that
determined from the spectroscopic abundance analyses -- the mean
abundance for the Paper II 63 star sample is $\langle$[Fe/H]$\rangle$
= --1.59, with dispersion 0.33, while for that of \citet{koch06} the
corresponding values are --1.91 and 0.28 using the \citet{zinn84}
calibration, and --1.73 and 0.35 using that of \citet{carretta97}.  In
summary, the upper RGB appears considerably tighter than one would
expect from the observed [Fe/H] values.

The obvious and simplest explanation of this conundrum lies in a
degeneracy of compensating variations of the other two parameters that
affect RGB morphology -- age and {\alphafe} (e.g. \citealp{monelli14,
  vandenberg15}).  A range in {\alphafe} values is a basic
characteristic within individual dSph, and \citet{venn12} have
demonstrated the existence of this effect in Carina. As
\citet{monelli14} and \citet{vandenberg15} have shown, the tightness
of the RGB could be explained by anti-correlations between [Fe/H] and
age and/or between [Fe/H] and {\alphafe}, such that the RGB sequences
of the various populations have an anti-correlation between [Fe/H], on
the one hand, and age and/or {\alphafe}, on the other, which conspire
to produce RGBs that closely overlap.  In Section~\ref{sec:syncmd} we
shall examine the role of both age and {\alphafe}, in addition to
[Fe/H], in seeking to decode Carina's observed CMD.

\subsection{A Comparison of Carina with Galactic Globular Clusters}\label{sec:ggc}

All high-resolution chemical abundance analyses of Carina red giants
to date report a large range in iron abundance, of order 1~dex between
the extreme values (\citealp{shetrone03, koch08a, venn12, lemasle12,
    {fabrizio12}, fabrizio15}).  In our homogeneous analysis of 63 red
  giant branch stars (Paper II) we find a range --2.5 $<$ [Fe/H] $<$
  --0.5.

In Figure~\ref{fig:cargc} we now compare Carina's CMD with those of
three Galactic globular clusters -- M13 and M92 (data from
http://www.cadc-ccda.hia-iha.nrc-cnrc.gc.ca/en/community/STETSON/standards/),
both of which show no spread in iron abundance ({\ltsima} 0.03 dex),
with [Fe/H] $\sim$ --1.6 and --2.3 (e.g. \citealp{kraft97, sneden00}),
respectively, and {\wcen} (data from \citealp{bellini09}), which
presents a large abundance spread --2.0 {\ltsima} [Fe/H] $<$ --0.7 --
see \citet{johnson10}.  To focus the discussion, at the top of each of
the four panels in the figure we have superimposed a box designed to
encompass the observations of the Carina upper RGB, which will appear
in several of the figures that follow.

We also superimpose \citet{dotter08} isochrones appropriate to M13 and
M92 on the figure, having parameters [Fe/H]/{\alphafe}/Age =
--1.6/0.4/13.8 and --2.2/0.4/13.8 respectively, together with distance
moduli and reddenings from \citet{harris96}(2010 edition)
(http://physwww.physics.mcmaster.ca/~harris/mwgc.dat), in panels (c)
and (b).  These parameters agree well with the literature values of
[Fe/H] = --1.53 and --2.31, respectively, of the Harris compilation;
with {\alphafe} $\sim$ 0.3 -- 0.5 (1D, LTE model-atmosphere modelling)
as summarized by \citet[Section 4.1]{sneden00} for globular clusters;
[O/Fe] = 0.5 for halo stars (3D, NLTE modelling) down to [Fe/H]
$\sim-2.5$ \citep{amarsi15}; and [Ca/Fe] = 0.3 (1D, LTE modelling)
reported in Paper II for Galactic halo stars; and the ages of 13.0 and
13.25 Gyr reported by \citet{dotter10}.  The differences between the
values we have adopted and those in the literature suggest that the
\citet{dotter08} isochrones can be fit to globular clusters to within
$\Delta$[Fe/H] = 0.05 -- 0.10, $\Delta${\alphafe} = 0.05 -- 0.10, and
$\Delta$Age = 0.7~Gyr.  One should also bear in mind the caveat of
\citet[Section 3.1]{vandenberg15} that the transformation from the
physical parameters (e.g., effective temperatures) obtained from
stellar evolution simulations to observed colors can lead to small
fitting errors between observation and theory.

The M13 and M92 isochrones are also superimposed on the Carina data in
panel (a).  Finally, the isochrone for M13 is superimposed on the data
for {\wcen} in panel (d).  The fit of the M13 isochrone to Carina on
the upper giant branch is very poor, where there is little, if any,
overlap between the RGB of M13 with that of Carina -- in spite of the
fact that the mean value for Carina of [Fe/H] = --1.6 is essentially
the same as that of M13.  This inconsistency graphically demonstrates
the need for lower ages and/or lower {\alphafe} values in Carina, as
discussed in the previous section.

Using the lower small boxes superimposed on the RGBs of M13, M92, and
{\wcen} in Figure~\ref{fig:cargc}, which cover the range 0.0 $<$ {\mv}
$<$ 1.0, we find color dispersions $\sigma$({\bi}) = 0.028, 0.025, and
0.077{\footnote{The widths of the boxes were chosen, by eye, to be
  sufficient to encompass the full color extent of the RGBs.},
respectively, for these clusters, consistent with the small, if any,
[Fe/H] and age spreads in M13 and M92, and the large abundance and
variously reported age spreads in {\wcen} (e.g., \citealp{joo13},
\citealp{villanova14}).

\subsection{Implications for [Fe/H] Abundances Determined with the Calcium~II -- Triplet Method}\label{sec:cat}

Analysis of the Ca II infrared triplet (at 8600~{\AA}) has provided a
powerful method for the determination of the [Fe/H] abundances of the
Galaxy's globular clusters and satellite dwarf spheroidal galaxies
(dSph) \citep{armandroff91}. There is, however, an important caveat
concerning the resulting dSph abundances, in particular the results
for Carina.  When the method of [Fe/H] abundance determination for
dSph systems is based on a calibration of the strength of the Ca II
features as a function of the height a red giant lies above the
horizontal branch in the CMDs of Galactic globular clusters (with
their mono-modal morphologies) (for Carina see \citealp {koch06}), on
the one hand, and the multiple-sequence CMD morphology and very
different evolutionary history of the dSph, on the other, one should
be concerned about the appropriateness of the calibration for Carina,
which has the most complicated CMD morphology of the Galactic dSph for
which data of sufficient quality are available\footnote{For comparison
  see \citet[Sculptor]{deboer11}, \citet[Sextans]{okamoto08b},
  \citet[Ursa Minoris]{okamoto08a}, and \citet[Draco]{grillmair98}}.

Specifically, the abundance of an individual red giant is given by an
equation of the form [Fe/H] = -- 2.706 + 0.326W$^{\prime}$, where
W$^{\prime}$ = W$_{\rm 8542}$ + W$_{\rm 8662}$ + 0.619($V$ -- {\vhb})
\citep{armandroff91}, and W$_{\rm 8542}$ and W$_{\rm 8662}$ are Ca II
equivalent widths and $V$ and {\vhb} refer to the magnitude of the
star and the horizontal branch having the same [Fe/H],
respectively. For Carina, \citet{koch06} adopt [Fe/H] = --2.77 +
0.38W$^{\prime}$, and an error in abundance of $\Delta$[Fe/H] =
0.21$\Delta(V$ -- {\vhb}), and report that an uncertainty in {\vhb} of
+0.4 mag leads to an error of --0.07 dex.  A further important
uncertainty may also be inferred from our Figure~\ref{fig:cargc}.  For
example, at [Fe/H] = --1.6, the data for M13 demonstrate a different
magnitude between globular clusters and Carina, such that at {\bio} =
2.3, $\Delta${\mv} = {\mv}$_{\rm Carina}$ -- {\mv}$_{\rm GC}$ =
--0.4~mag.  In the \citet{armandroff91} calibration, this corresponds
to an abundance difference $\Delta$[Fe/H] = $\sim$ --0.1~dex.  As
discussed above, this effect is driven by differences in the age and
{\alphafe} distributions between Carina and the Galactic globular
clusters.  A further shortcoming of the use of Galactic globular
clusters as calibrators is that the [Ca/Fe] distributions differ
significantly between them and the dSph.  For example, at [Fe/H] =
--1.5, the difference between Carina and the halo is $\Delta$[Ca/Fe] =
[Ca/Fe]$_{\rm dSph}$ -- [Ca/Fe]$_{\rm Halo}$ $\sim$ --0.2 (see e.g.,
\citealp{tolstoy09}). One should not be surprised then to find errors
in [Fe/H] of $\sim -0.3$ dex in this formalism.

We investigate this further by comparing the (Ca II triplet based)
[Fe/H] abundances of \citet{koch06} with the high-resolution
spectroscopic values of Paper II.  In Figure~\ref{fig:norkoch}, the
leftmost panel, (a), presents the CMD for the 35 stars in common
between the two investigations; the middle panel, (b), compares the
[Fe/H] values from the two investigations; and the rightmost panel,
(c), plots the distance, $\Delta${\bio} (= {\bio} -- (\bi)$_{\rm
  RGB}$), a star falls from the fiducial RGB in the CMD in panel (a)
as a function of the abundance difference [Fe/H]$_{\rm Koch06}$ --
[Fe/H]$_{\rm Paper II}$.  In panel (b) we also plot representative
estimated abundance error bars, both of which are $\sim$~0.13~dex in
size.  In panels (b) and (c) the blue symbols represent outliers
(defined in (c) where they comprise some 15\% of the sample).
Consideration of panel (b) shows that $\sim$ 10\% of the sample lie
further than $\sim$~3$\sigma$ from the 1 -- 1 line.  We conclude that
further consideration of this method is needed.
 
We would argue that a more appropriate approach to determining [Fe/H]
from measurements of the Ca II triplet would be a theoretical one
using a model atmosphere analysis together with atmospheric parameters
based on a star's position in the CMD (see e.g., \citealp{norris08,
  starkenburg10}). That said, however, we note that even the results
for Carina of \citet[their Section 4.1]{starkenburg10} may be open to
question, insofar as they report ``We find that for a range of old
ages, between 8 and 15 Gyr, the exact choice of the isochrone age does
not significantly affect our results.''  As we shall discuss in our
Section~\ref{sec:upperrgb}, we find that some 75\% of the Carina's RGB
stars have ages below 8 Gyr, suggesting that extrapolation in the
determination of [Fe/H] may have occurred for these objects.  Given
that a significant percentage of Carina stars appear to be younger
than the ages used in the \citet{starkenburg10} analysis, we suggest
that this calibration could be revisited.

In summary, Carina [Fe/H] abundances determined from analysis of the
Ca II infrared triplet use calibrations of questionable validity in
the present context, and should be treated with caution. 

\section{SYNTHETIC COLOR MAGNITUDE DIAGRAMS}\label{sec:cmd}

We seek to interpret the very high quality $BVI$ photometry of Carina
described by \citet{stetson11} and made available by P.B. Stetson
(2014, private communication) by comparing its ({\mv}, \bio) CMD
against synthetic color-magnitude diagrams based on the isochrones of
\citet{dotter08}.  In our analysis we adopt a set of \citet{dotter08}
isochrones defined by: [Fe/H] = --2.45 (0.05) --0.70, {\alphafe} =
--0.20 (0.10) 0.40, age = 0.8 (0.2) 15.0 (Gyr), and helium abundance Y
= 0.245 + 1.5*Z (where Z is the mass fraction of elements heavier than
helium){\footnote{We realize that substantial helium variations exist
    among the populations within several Galactic globular clusters
    such as, for example, $\omega$ Cen for which $\Delta$Y $\sim$ 0.12
    (\citealp{norris04, piotto05}). To our knowledge there has been no
    suggestion that large helium variations exist in the dSph systems,
    and for economy of hypothesis we assume that none does.}}, together with a
Salpeter mass function.

\subsection{The Four Basic Populations} \label{sec:4p}

Figure~\ref{fig:cariso} presents the ({\mv}, \bio) CMD where the lower
panel covers the magnitude range --3.0 $<$ {\mv} $<$ 4.4, and the upper
panel covers the upper giant branch in the range --3.0 $<$ {\mv} $<$
0.0.  The leftmost panels contain the observational data, together
with representative error bars (to the right) determined by averaging
the (internal) photometric errors in $V$ and {\bi} available in the
Stetson catalog.  These errors are also presented in
Table~\ref{tab:errors}.  As outlined in Section~\ref{sec:decoded}, we
seek to model Carina in terms of four ``basic'' populations, to explain
four of the observationally prominent features defined in
Section~\ref{sec:carina_cmd}.  In the figure we use four boxes to
define regions of the CMD that each contain stars from only one of
these populations.  These are the four boxes fainter than {\mv} = 1.0,
and the faintest and reddest box contains the first population, with a
monotonic progression to the brightest and bluest box which contains
the fourth.  The fifth and brightest box contains stars on the upper
giant branch encompassing the 63 star RGB star sample for which
chemical abundances are available based on high-resolution spectra to
be presented in Paper II.

To determine the relevant stellar populations, we adopt the isochrones
illustrated in the CMDs in the lower panels of
Figure~\ref{fig:cariso}. Here, and in what follows, we consider two
cases for {\alphafe}: in the middle panels we present isochrones
having constant {\alphafe} = 0.1, while in the rightmost panels we
permit {\alphafe} to vary over the range --0.20 to +0.20.  In the lower
middle and right panels the legends contain the basic
[Fe/H]/{\alphafe}/Age parameters of the four populations.  As will be
described in more detail below, for each of the faintest three
populations in the lower panels (the so-called first, second, and
third populations), we adopt one basic isochrone chosen to pass in a
representative manner through the stars in the population isolating
boxes described above. We shall return to the fourth population below.

\subsubsection{The First Three Populations}\label{sec:oldpop}

We illustrate our procedure to determine ages and abundances for the
case of constant {\alphafe} = 0.1, describing the choice of the
isochrone for each of the three populations in the defining boxes in
the lower, middle panel.  Given the tightness of the RGB, we firstly
require that each isochrone passes as closely as possible through the
point ({\mv}, {\bio}) = (0.5, 1.62) on the RGB (as mandated by the
observations); then for the first population, that it should also pass
through ({\mv}, {\bio}) = (3.25, 1.20); and for the second and third
populations pass through ({\mv}, {\bio}) = (2.72, 0.97) and (2.50,
0.42), on their respective SGB and main-sequence stub, as seen in the
observed CMD.  The results of this process are presented in
Figure~\ref{fig:loci}.  Here the relatively horizontal lines represent
the (age, [Fe/H]) -- loci of isochrones that pass through the boxes
defining the SGBs and main-sequence stub of the first, second, and
third populations, while the more vertical line crossing these three
presents the locus for isochrones that pass through the RGB at ({\mv},
{\bio}) = (0.5, 1.62). 

The intersections in Figure~\ref{fig:loci} provide the ages and
abundances of the isochrones of the first three populations in the
middle panels of Figure~\ref{fig:cariso}, and presented in columns (2)
-- (4) of Table~\ref{tab:pops}.  For the variable {\alphafe} case, we
chose {\alphafe} values of 0.2, 0.1, and 0.0 for the first through
third populations, and for each of these the [Fe/H] and age values
were determined in a similar manner to that described above.  The
adopted values of [Fe/H]/{\alphafe}/Age for this case are presented in
the rightmost lower panel of Figure~\ref{fig:cariso} and in columns
(6) -- (8) of Table~\ref{tab:pops}.

\subsubsection{The Fourth Population}\label{sec:fourthpop}

The simulation of the ephemeral, putative main-sequence, fourth
population at ({\mv}, {\bio}) = (2.0, 0.0) -- (3.0, 0.2) is more
challenging.  Here there is an obviously larger observed spread in
color at a given magnitude ($\Delta${\bio}~$\sim$~0.2 -- 0.3 at {\mv}
= 2.0) than might be expected from errors in measurement of 0.025 dex
(see, e.g., Table~\ref{tab:errors}) and the dispersion of the Carina RGB
of $\sigma${\bio}~$\sim$~0.04, at the same magnitude.  Further, the
main-sequence feature exhibits a tantalizing suggestion of structure
(in the range 2.0 $<$ {\mv} $<$ 3.0).  One might wonder if more than
one sub-population exists within our so-called fourth population.  As we
shall see we are somewhat limited in our investigation by the fact
that the minimum age in the set of isochrones we have chosen to use is
0.8 Gyr.  With this background, we have adopted a single ``basic''
isochrone for the fourth population in the middle panels of
Figure~\ref{fig:cariso} having parameters [Fe/H]/{\alphafe}/Age =
--1.2/0.1/1.4 for the constant {\alphafe} case, while in the rightmost
panels we assume --1.1/--0.2/1.4 for the variable {\alphafe}
case\footnote{We note for completeness that our choice of the range
  for the variable case ({\alphafe} = --0.2 to +0.2) for the four
  populations was guided by what is known about Carina's observed values
  (e.g., \citealp{venn12} and Paper II).}.  We shall address the nature of
the fourth population further in Section~\ref{sec:spreads}.

\subsection{Synthetic CMDs of the Basic Populations}{\label{sec:syncmd}}

\subsubsection{First Approximation}{\label{sec:approx}}

First-order synthetic CMDs are presented in Figure~\ref{fig:carsyn1},
where the general layout is the same as in Figure~\ref{fig:cariso}.
As before, the observational data for the total Carina sample are
presented in the bottom left panel.  To its right are synthetic CMDs
based on the four ``basic'' populations shown in
Figure~\ref{fig:cariso} for fixed {\alphafe} (middle panel) and
variable {\alphafe} (right panel). In Figures~\ref{fig:cariso} and
~\ref{fig:carsyn1}, and throughout this paper, red, green, blue, and
magenta are used to refer to the first through fourth populations,
respectively.

To produce the synthetic CMD, we drew magnitudes and colors at random
from the \citet{dotter08} isochrones adopting the Salpeter
mass-function.  To these we added random observational uncertainties,
for which we have presented representative values in
Table~\ref{tab:errors}.  The lower panels of Figure~\ref{fig:carsyn1}
contain the five boxes described in Figure~\ref{fig:cariso}, which
were chosen to isolate important representative populations in the CMD
(the faintest four of which isolate samples unique to the four basic
Carina populations).  Also presented in the legend of the lower
leftmost panel are the numbers of stars observed in the five
sub-samples, corrected for background interlopers by using a nearby
field some 25{$\arcmin$} from the center of Carina\footnote{We note
  for completeness that we normalized the counts in the background
  field to those of the program field in several sub-regions in the
  CMD well away from the Carina principal sequences.}.  The
corrections to the number of stars in the boxes isolating the first
through fourth populations were 204, 115, 5, and 15, respectively, and
72 for the box on the upper RGB.

In preparing the synthetic CMD we chose stars at random for each of
the four basic populations in turn, until we obtained the observed
number of stars in the relevant box in Figure~\ref{fig:cariso}.
Finally, we simulated field non-members by superimposing on the
diagram the data from the nearby observed field mentioned above, which
we added to the synthetic catalog.  The synthetic CMDs are presented
in the middle (constant {\alphafe}) and rightmost (variable
{\alphafe}) panels, where the star counts in the lower middle and
lower right panels (corrected for background) were determined from the
synthetic CMDs.  These numbers differ slightly from the targets
presented in the observed CMD (left panel), due to contamination by
other populations.  In the lower panels of the synthetic CMDs, the
fifth star count corresponds to the box on the upper RGB, and
represents the (predicted) cumulative numbers for the four synthetic
populations.

Having inspected Figure~\ref{fig:carsyn1}, the reader will immediately
realize that we have made no attempt to simulate post-RGB evolution.
Insofar as the horizontal branch evolutionary phase is not essential
to the problem we are addressing, this is not a serious concern.  Of
some importance, however, is our neglect of the AGB.  As discussed in
Section~\ref{sec:tightness}, above {\mv} $\simlt~-1.0$, the AGB
constitutes some 20 -- 30~\% of stars in the Galactic globular
clusters. This is, of course, not the complete story.  As the presence
of some nine carbon stars in Carina (\citealp{azzopardi86} and
references therein) attests, one should also consider the contribution
of AGB stars from intermediate-age populations. To address this issue,
we estimated the fraction of AGB on the giant branch using the results
of \citet{girardi00}.  For their isochrone having Z = 0.0004 ([Fe/H]
$\sim-1.6$) and age = 7.1~Gyr, we estimate that the AGB represents some
20\% of red giant stars in the magnitude range --2.6 $<$ {\mv} $<$
--0.7. We shall return to the question of the absence of AGB from our
synthetic CMDs in Section~\ref{sec:upperrgb}, when we compare the
number of stars on the upper giant branch of Carina's observed and
synthetic CMDs.

Even so, the astute reader will also note that at best there is only
generic agreement between observation and theory in
Figure~\ref{fig:carsyn1}: the model produces sharper sequences than
are observed; and, indeed, there are three clearly predicted subgiant
branches, while only two are observed.  We would have to agree.  The
problem, of course, is that our model of the four putative ``basic''
populations, each with its unique combination of [Fe/H], {\alphafe},
and age, is a gross over-simplification, in the sense that one might
expect a spread in age and abundances within each of the basic
populations.

\subsubsection{Abundance and Age Spreads within the Four Populations}{\label{sec:spreads}}

The existence of clear sequences (except perhaps for the fourth
population) in the observed CMD leads to the expectation that the
sizes of the spreads of physical parameters within each of the
populations are relatively small.  We shall see that moderate age
spreads within a given population are more efficient in Carina in
producing morphological changes than those in abundance. Previous
estimates of age spreads were made in the earliest studies:
\citet{smecker96} reported that ``bursts'' and quiescent phases lasted
for {\gtsima}1 Gyr, while \citet{hurley98} found that the major
intermediate age component may have lasted as long as 2 Gyr.  We now
consider the implications of these spreads within the first, second,
and third populations.

\subsubsubsection{The First Three Populations}

We begin with the tightness of the RGB.  As discussed in
Section~\ref{sec:tightness}, in the magnitude range 0.0 $<$ {\mv} $<$
1.0 the RGB has an RMS dispersion of $\sigma$({\bio}) = 0.038 mag
about the quadratic best fit.  If we consider the basic
\citet{dotter08} isochrones adopted for the oldest three synthetic
Carina populations in the middle panel of Figure~\ref{fig:carsyn1}
([Fe/H]/{\alphafe}/Age = --1.2/0.1/3.4, --1.5/0.1/7.0, and
--1.85/0.1/13.2), for $\sigma$({\bio}) = 0.038 mag the corresponding
dispersions in [Fe/H] in the three populations that could produce the
observed dispersion in color are 0.1 -- 0.2 dex.  If the three
populations do not exactly overlap in color in the CMD in the above
magnitude range, then in order to reproduce the observed spread in
{\bio}, the inferred individual dispersions in [Fe/H] would be even
smaller.  Variations in age within the three populations should also
be considered.  Dispersion in age of 1 Gyr at {\mv} = 0.5 on the RGB
would lead to dispersions of $\sigma$({\bio})~$\sim$~0.03, 0.02, and
0.01 (age increasing), respectively, which would reduce the inferred
sufficient dispersion in [Fe/H].  Variations in {\alphafe} cause only
small changes in $B-I$, and given this relative insensitivity, we
choose not to consider the effect of a spread in {\alphafe} within the
diffused populations in what follows.  From a different viewpoint, the
limits that the tightness of the RGB places on the range in age within
the three populations are, roughly speaking, $\sim$~1 -- 4 Gyr.

In the absence of obvious strong theoretical constraints on the form
of the abundance and age distributions within individual populations
and their potential correlations (and for simplicity and convenience)
we assumed that each of the oldest three populations has small spreads
in [Fe/H] and age, which we represent by tophat distributions, about
the basic values adopted in Section~\ref{sec:4p}.

Figure~\ref{fig:carwobble1} shows the result when small spreads of
fullwidth 0.1~dex and 1.2 -- 2.8~Gyr are applied to the basic [Fe/H]
and age values, respectively, of the oldest three populations in
Figure~\ref{fig:carsyn1}.  The improvement to the simulations is
clear.  Of particular interest is that the third population SGB, at
{\mv} = 2.2, which is clearly present and was commented upon in
discussing Figure~\ref{fig:carsyn1}, is no longer obvious when spreads
in [Fe/H] and ages are taken into account.  Figure~\ref{fig:sgb} further
examines the role of the spreads of [Fe/H] and age (principally the
latter) in diffusing each of the sequences within the CMD.  In the
figure, the left panel refers to the observations, the middle to the
simulations in Figure~\ref{fig:carsyn1} ({\alphafe} = 0.1) smoothed
only by observational errors, and the right to those in
Figure~\ref{fig:carwobble1} ({\alphafe} = 0.1) smoothed by both
observational errors and the spreads in [Fe/H] and age adopted in that
figure.  Here one sees clearly the smearing out effect on the third
population.  

Also superimposed on these panels is a grid that facilitates
estimation of the absolute magnitude dispersions on the individual
SGBs. The upper panels in the figure plot the number of stars in each
of the 35 boxes of the grid, as a function of box number (Z),
beginning with the fainter stars.  The improvement of the simulations
of the SGBs when the small dispersions in [Fe/H] and age are included
is evident.  A further check on the dispersing effect of the age
spread for the third population is to count the stars within an
appropriate section of the grid in Figure~\ref{fig:sgb}.  We choose to
do this for the bottom 12 boxes of the grid (i.e. over the range 1.7
{\ltsima} {\mv} {\ltsima} 2.3), where we find that the
background-subtracted number of stars, 220, in the observed Carina CMD
on the left, compares satisfactorily with the number, $\sim$185, found
in each of the two synthetic CMDs in the figure.

\subsubsubsection{The Fourth Population}{\label{sec:fourth}}

A challenge for the fourth population is that the color range is
surprisingly large, say, $\Delta${\bio}~$\sim$~0.25 at {\mv} = 2.0,
not only in comparison with the observational errors, $\sim0.04$~dex,
but also with the observed dispersion of the RGB of
$\sigma${\bio}~$\sim0.04$ at that magnitude (see
Section~\ref{sec:fourthpop}).  In order to discuss the difficulties,
in Figure~\ref{fig:carsyn1} we represented the fourth population by
the basic isochrones [Fe/H]/{\alphafe}/Age = --1.20/0.10/1.4 and
--1.10/--0.2/1.4 for the constant and variable {\alphafe} cases,
respectively, and then in Figure~\ref{fig:carwobble1} broadened the
population with tophat filters having fullwidths in [Fe/H] and age of
0.1~dex and 1.2~Gyr, respectively, similar to our procedure for the
first three populations.  Here the color dispersion of the fourth
population in the synthetic CMD (variable {\alphafe} case) in the
magnitude range {\mv} = 2.0 -- 2.4 is $\sigma${\bio}~$\sim0.055$
$\pm$~0.005, still significantly narrower that the Carina
observations, for which the corresponding dispersion is
$\sigma${\bio}~$\sim0.088$ $\pm$~0.008.

A basic problem of the simulation is that the synthetic main sequence
of the fourth population does not reach as blue as the observations.
One limitation of the model might be that it uses isochrones reaching
down to only an age of 0.8 Gyr, which is at the low limit of the age
range of the stellar isochrones adopted in this work -- that is, we
cannot reach ages lower than 0.8 Gyr in our simulations.  In
Figure~\ref{fig:carwobble1}, with a basic synthetic age of 1.4 Gyr and
an age spread of 1.2 Gyr about that value we have reached that limit.
Younger isochrones are needed to address this problem.  Further, in
the variable {\alphafe} case, in the rightmost panels of
Figure~\ref{fig:carwobble1} we have essentially reached the lower
limit of {\alphafe} = --0.2 indicated by the observations (Paper II,
Figure 20), and the lower limit available in the \citet{dotter08}
isochrones.  In principle, we could force the simulations of the main
sequence bluer by decreasing the metallicity.  This, however, seems
intuitively inappropriate given the monotonically increasing value of
[Fe/H] as one progresses from first to third populations; that is, at
first thought one might expect a higher value than [Fe/H] = --1.1 to
--1.2 for the fourth population.  We shall revisit this problem in
Section~\ref{sec:raddis}.

Are there other possibilities?  We noted in Section~\ref{sec:decoded}
that we thought it unlikely that the fourth population was actually
part of Carina's horizontal branch.  Perhaps the HB provides a partial
explanation of the observations and we are observing an admixture of a
young main sequence and a very faint blue horizontal branch?

A second, and in our view more likely candidate, is that we are
observing a young main sequence that has been broadened by stellar
rotation.  While there is no counterpart of the young, metal-poor main
sequence fourth population in the Milky Way, there is a roughly
similar component in the intermediate age populous clusters in the
Magellanic Clouds.  As first pointed out by \citet{mackey08}, there is
a much larger spread in the color of main-sequence turnoff stars in
these clusters than can be explained by the photometric uncertainties,
which \citet{bastian09}, and \citet{brandt15} have discussed in terms
of structural changes in stars with rotation rates that are 20 -- 50
\% of the critical rotation rate for breakup.  See, for example,
Figure 1 of \citet{brandt15}.  

A third possibility, advocated recently by \citet{santana16}, is that
our fourth population comprises blue stragglers of their intermediate
age population (our second and third populations).  While we are
unable to rule out this possibility, we note that their argument holds
only at the ``1 standard deviation difference'' level (see their
Section 5.1, Figure 12)\footnote{A further complication with this
  suggestion, in our opinion, is that Carina's principal population is
  considerably younger than those of the calibrating systems adopted
  by \citet{santana16} (Sculptor, Draco, Sextans, and Ursa Minor (see
  e.g. \citealp{weisz14} and \citealp{lee09})), calling into question
  the reliability of their basic calibration).}.

Our view is that a resolution of the nature of the fourth population
will be solved only when spectra of sufficient resolution become
available to distinguish between these putative HB stars, blue
stragglers, and rotating and non-rotating, metal-poor, main-sequence
stars.

\subsubsection{Radial Distributions}\label{sec:raddis}

Further potential insight into the above discussion is provided by
consideration of the radial distributions of the four populations,
which we present in Figure~\ref{fig:raddis}, where we plot the
cumulative distributions of the populations as a function of elliptic
radius, r$_{\rm ellip}$.  The first thing to note is that the
concentrations of the first three populations increase as one goes
from the (oldest) first population to the (more recent) third one.  In
stark contradistinction, however, the concentration of the (putatively
youngest) fourth population is less than that of any of the older
ones.  The simplest explanation of this is that there is little if any
direct physical connection between the fourth population and the other
three.  Given the dependences of {\bi} on [Fe/H], {\alphafe}, and age
presented above in Section~\ref{sec:fourth}, we conjecture that the
metallicity of the fourth population is significantly lower than our
first suggestion of [Fe/H] $\sim$ --1.1, as postulated there.

In Figure~\ref{fig:carwobble2}, for heuristic purposes, we present
synthetic CMDs similar in all respects to those in
Figure~\ref{fig:carwobble1} except that the metallicities of the
fourth population are lower by $\Delta$[Fe/H] =~0.3 dex than in the former.
There are two interesting points of difference.  First, the main
sequence of the fourth population is bluer by $\sim$ 0.10 in
Figure~\ref{fig:carwobble2} than in Figure~\ref{fig:carwobble1},
improving a little the agreement between observation and model. (We
note for completeness that the improvement could probably be increased
further if we were to have isochrones that reached to ages lower than
0.8~Gyr.) Second, on the red giant branches in the upper panels of the
middle and rightmost columns of Figure~\ref{fig:carwobble2}, one sees
a handful of stars from the fourth population bluer by $\Delta${\bi}
$\sim$~0.4 mag than the RGB at the same {\mv} (and also that of their
counterparts in Figure~\ref{fig:carwobble1}).

The reader may recall that in Section~\ref{sec:tightness} we pointed
out a handful of Carina radial-velocity members in the catalogs of
\citet{koch06} and \citet{walker09} that fell significantly blueward
of the large majority of Carina's RGB.  It is tempting to further
conjecture that these observed objects can be identified with stars in
our synthetic fourth population.

We realize that the implications of our conjectures are considerable,
involving as it does the implicit suggestion that the fourth
population has a very different spatial distribution than the third
population in an environment that has remained relatively unenriched.
We await further targeted observations to see if these suggestions
survive rigorous testing.

To conclude this discussion, we note that \citet{kordopatis16} have
reported that Carina has three populations of red giants which they
designate ``metal-poor'' ($\langle$[M/H]$\rangle$ = --2.4),
``intermediate-metallicity'' ($\langle$[M/H]$\rangle$ = --1.84) , and
``metal-richer'' ($\langle$[M/H]$\rangle$ = --1.0).  They too have
found that their most metal-rich component is less centrally
concentrated than the more metal-poor populations.  Given the
different approaches of the two investigations (and not least the
resulting different number of components), however, some caution
should be exercised in claiming that both investigations have observed
the same phenomenon.  That said, we note that the \citet{kordopatis16}
``metal-richer'' component comprises 20\% (by number) of the red
giants studied, while our fourth population represents 6\% of the
upper RGB (see Section~\ref{sec:upperrgb}).

Given our uncertainty about the metallicity of the fourth population,
in what follows we shall arbitrarily present results for that
component for the more metal-rich cases ([Fe/H] = --1.1 (constant
{\alphafe}) and [Fe/H] = --1.2 (variable {\alphafe}).  Insofar as the
fourth component is a significantly minority component, this choice
will have only a small effect on the results that will be presented.

\subsubsection{Comparison of the Ages of This Work with Those of Earlier Investigations}

It is of some interest to compare the results of the present work with
those of others in the literature.  Table~\ref{tab:litages} presents
ages from investigations that have reported multiple populations in
Carina, where the data are given to the precision reported by those
authors.  Our results are presented in the final row.  Column (1)
gives the authors, while columns (2) -- (5) contain the ages (or
mean/median values when appropriate) of the populations, in descending
order.

Inspection of the table shows that different authors report different
numbers of components.  This is perhaps not unexpected, insofar as the
data for different investigations are of different quality and
analyzed in different ways, as a result of which some investigations
do not resolve neighboring components or those that are of only minor
significance.

What are the essential/common results that one might take from the table?

\subsubsubsection{The Oldest Population}

Inspection of Figure~\ref{fig:loci} and Table~\ref{tab:pops} of the
present work leads to the conclusion that the mean age of the oldest
population in Carina's CMD is not significantly less than 13~Gyr,
considerably more restrictive than the comment by \citet{deboer14}
that ``Different studies agree in general that the intermediate-age
episode took place somewhere between 3 and 8 Gyr ago and the older
episode $>$~8 Gyr ago, but the exact age and duration of the episodes
still remains uncertain''.  Our 13~Gyr is also inconsistent with the
7.5 Gyr of \citet{hernandez00}. That said, in the light of the six
results in Table~\ref{tab:litages} favoring an age greater than 11
Gyr, it seems reasonable to suggest that the median age of the first
population lies in the range 11 -- 13 Gyr.

\subsubsubsection{How Many Intermediate-Age (2 --7 Gyr) Populations
  Are There?}

If one sets aside works that present only a (large) age range for the
intermediate age population(s), as opposed to distinct components
within the adopted limits, there seems to be a reasonable case for two
populations in the range 2 -- 7 Gyrs.

\subsubsubsection{The Youngest Population?}

Four of the nine investigations in Table~\ref{tab:litages}, and in
particular those based on recent very high quality faint CMDs, report
a very young main sequence component that they have interpreted in
terms of a $\sim$1 Gyr metal-poor component.  This interpretation has
been challenged by \citet{santana16}, who suggest that this putatively
young component comprises the blue stragglers of the intermediate-age
(2 -- 7 Gyr) populations. While we are not persuaded to this position
(see our comments in Section~\ref{sec:fourth}), further work is needed
to resolve the issue.

\subsubsection{The Masses of the Four Populations}

One may use the above formalism to directly estimate the masses of the
four populations within the elliptical radius, r$_{\rm ellip}$ =
13.1{$\arcmin$}, under investigation here.  Summation over the
synthetic CMD brighter than {\mv} = 4.5 for the constant {\alphafe}
case in Figure~\ref{fig:carsyn1} yields masses of 2820, 6850, 7150,
and 2090 M$_{\odot}$ for the first through fourth populations,
respectively.  Expressed as fractions of the total mass these
correspond to 0.15, 0.36, 0.38, 0.11\footnote{We realize that the use
  of Salpeter or similar power-law Initial Mass Functions is at odds
  with what is generally found in the Galaxy for very-low-mass stars,
  where the IMF is shallower at the low-mass end (e.g.,
  \citealp{kroupa01}).  \citet{wyse02} have reported a similar result
  for the UMi dSph.  Hence our total mass estimates may be too high.
  We adopt the Salpeter formalism for consistency with the isochrones
  we have adopted, and present the above results principally for
  heuristic purposes.}.

As noted in Section~\ref{sec:approx}, our synthetic CMDs were
determined by adopting a Salpeter mass function, n(m) =
n$_{0}{\times}$m$^{-2.35}$, where the number of stars n(m) formed
between masses m and m+dm is n(m)dm, and n$_{0}$ is a constant.  It
follows that the number of stars, N, in the mass range m$_{1}$ to
m$_{2}$ is N = n$_{0}$/1.35$\times$(m$_{1}^{-1.35}$ --
m$_{2}^{-1.35}$), while the total mass of the population in the same
mass range is M = n$_{0}$/0.35$\times$(m$_{1}^{-0.35}$ --
m$_{2}^{-0.35}$). For each of the four populations we used the stars
in the synthetic CMD between appropriately chosen mass limits
pertinent to each of the four boxes we adopted to isolate only stars
from a given population in order to determine the value of the
constant n$_{0}$, and then used the mass equation to determine the
total mass of the population (at formation) between the lower and
upper mass limits of Carina, for which we adopt 0.1 and 60M$_{\odot}$.
The resulting total masses of the four populations at their times of
formation are then 2.12$\times10^{5}$, 2.48$\times10^{5}$,
1.42$\times10^{5}$, and 0.24$\times10^{5}$ M$_{\odot}$ (with mass
fractions 0.34, 0.39, 0.23, and 0.04)\footnote{To the reader who is
  surprised by the apparent disagreement between the relative masses
  fractions of the four populations in the CMD compared with those
  over the whole of the assumed IMF, we note that for a given
  population the ratio is a strong function of the relevant mass range
  in the CMD over which the determination is made.}. These mass
fractions are also presented in column (5) of Table 3.  The coaddition
of these masses is 6.3$\times10^{5} $M$_{\odot}$, which provides some
insight into the total baryonic mass that initially formed stars.  It
represents the total initial baryonic mass associated with material
currently within the r$_{\rm ellip}$ = 13.1{$\arcmin$} at the various
epochs of formation.  We complete this section by noting that
\citet{deboer14} report that the ``total stellar mass formed over the
duration of star formation of Carina ... is 0.43 $\pm$ 0.05
$\times10^{6} $M$_{\odot}$ within the half-light radius of 250 pc''
(corresponding to 9.1{$\arcmin$} at a distance of 106 kpc).  One
should also bear in mind that our values may be overestimates.  In low
metallicity objects such as Carina, star formation must have been
highly inefficient, so large masses would very likely have been lost
from the system.

\section{THE POPULATION STRUCTURE OF THE UPPER RED GIANT BRANCH}\label{sec:upperrgb}

We now use the results of the synthetic CMD presented in
Figure~\ref{fig:carwobble1} to seek insight into the population
structure of Carina's upper giant branch.  The breakdown of the
populations is shown in Figure~\ref{fig:synrgb} for the variable
{\alphafe} case.  On the upper RGB the first -- fourth populations
comprise 14\%, 38\%, 42\%, and 6\% of the sample, respectively.

The total number of synthetic stars in the box on the upper giant
branch in Figure~\ref{fig:carwobble1} (119, variable {\alphafe} case),
is somewhat smaller than that observed (169).  We have argued in
earlier sections that the synthetic result should be increased by 20
-- 30\% to allow for AGB stars.  We would also note here that in the
observed Carina CMD in the magnitude range --1.0 $<$ {\mv} $<$ 0.0,
there appear to be commensurate numbers of likely AGB and First-GB
stars, for which N$_{\rm AGB}$/(N$_{\rm AGB}$+N$_{\rm First-GB}$)
$\sim$ 0.4.  Be that as it may, for a 25\% AGB component, the
predicted number would increase to 149, only 10\% lower than that
observed; and given the numbers, a result significant at only the
1.1$\sigma$ level.  Finally, we comment on the question of
completeness of the Carina CMD as one goes to faintest magnitudes.
While, to our knowledge, no estimates are available in the literature
for the present CMD, potential completeness information is available
from the deep (g, g--r) photometry of \citet[their Figure
  2]{santana16}, from which we estimate 5 -- 15\% incompleteness over
the magnitude range {\mv} = 2.5 -- 3.3 of our third through first
populations.  If this were relevant to the present investigation, it
would lead to a decrease of the same order in the predicted number of
stars in our upper giant branch box.

The distribution functions of [Fe/H], {\alphafe}, and age for the
synthetic CMD are presented as generalized histograms in
Figure~\ref{fig:moddistrib} for the variable {\alphafe} (upper panels)
and fixed {\alphafe} (lower panels) cases.  The results for the four
individual populations are shown in different colors as described in
the figure caption, while those of the total population have been
overplotted in black; Gaussian kernels of 0.15 dex, 0.10 dex, and 0.5
Gyr have been adopted for [Fe/H], {\alphafe}, and age, respectively.
(Here the smoothing kernels were chosen somewhat arbitrarily, to
clearly show the identity and roles of the populations.)  We make two
comments on the figure. First, as already noted, the bulk of Carina
lies at intermediate age.  Following our discussion in
Section~\ref{sec:cat} on the analysis by \citet{starkenburg10} of the
Ca II triplet to produce iron abundances, we specifically note here
that we find $\sim$75\% of the system is younger than 8 Gyr, casting
some uncertainty on the abundances of that analysis, insofar as it
assumed ages older than that limit.  We also comment that it has been
suggested that the treatment of age has little effect on [Fe/H]
abundances determined using the Ca II triplet, based on the work of
\citet{cole04} and \citet{battaglia08}.  Our concern is that while
empirical calibrations of the Ca II triplet are accurately determined
for the bulk of old systems within the Galaxy, it is not clear to us
that this is the case for a system as complicated in its population
structure as Carina.  For example, if one plots the Galactic globular
and old open clusters studied by \citet[their Table 1]{cole04},
together with the four populations of Carina in our
Table~\ref{tab:pops}, in the (Age, [Fe/H]) -- plane, one sees that the
second through fourth Carina populations lie far from the Galactic
objects.  That is, with ages of $\sim2 - 7$ Gyr and abundances [Fe/H]
$\sim$ --1.5 to --1.15, the Carina populations lie below the Galactic
clusters by 5 -- 10 Gyr, and at abundances lower by 0.8 -- 1.0 dex --
well away from the fundamental Galactic calibrations.

Second, while the differences between the two {\alphafe} cases are not
large for the [Fe/H] and age distributions, those for the {\alphafe}
distributions are significantly different, as might be expected.  We
shall return to this issue in Paper II (Section 10).

The reader will comment that the age Gaussian kernel of 0.5 Gyr in
Figure~\ref{fig:moddistrib} is far too small given the difficulty in
determining ages for stars on the RGB from a color-magnitude diagram.
A more realistic estimate might be 4.0 Gyr (see, e.g., Paper II),
which we adopt here in both upper and lower panels to determine the
age distributions in Figure~\ref{fig:modagedist} for the constant and
variable {\alphafe} cases.  The lack of sensitivity in both panels is
obvious.  We shall return to a comparison of this distribution with
that of the age estimates based on the analysis of the Carina upper
giant branch in Paper II.

\section{SUMMARY AND DISCUSSION}\label{sec:discussion}

\begin{itemize}

\item  

We have used the isochrones of \citet{dotter08} to produce synthetic
CMDs that provide a basic description of four epochs of star formation
in Carina, in terms of the [Fe/H], {\alphafe}, and ages of the four
populations, which have [Fe/H] = --1.85, --1.5, --1.2, and
$\sim$--1.15 and ages $\sim$~13, 7, 3.5, and $\sim$1.5 Gyr,
respectively, and two {\alphafe} cases.  In the first, {\alphafe} =
0.1 (constant {\alphafe}), while in the second {\alphafe} = 0.2, 0.1,
0.0, --0.2, respectively (variable {\alphafe}).  We refer to these as
the first through fourth populations (based on their ages) for which
details are presented in Table~\ref{tab:pops}.  The parameters for the
first three populations are strongly constrained by Carina's
well-defined subgiant and main-sequence features together with its
very tight RGB.  The nature of the fourth population is not so clear,
and future work is needed to determine whether a lower value of age and/or
[Fe/H], blue stragglers, or stellar rotation may also play a role
in its position in the CMD.

\item  

A strong (age, metallicity) -- relationship emerges from this
analysis, in particular for the first through third populations, which
is based only on our fitting of four well-defined features in the CMD
-- distinct subgiant branches of the first and second populations, the
conspicuous upper main-sequence stub of the third population, and the
small color dispersion observed on the RGB.  We shall return to the
(age, metallicity) -- relationship in Paper II (Section 9).

\item 

As part of the process, we made first order estimates of the spreads
in [Fe/H] and age for the four basic populations, which are required
to explain the spreads in color and magnitude within the principal
sequences of the observed CMD.  In the absence of other constraints,
we assumed that the spreads in these parameters may be described by a
tophat function, in which the basic population is spread evenly over
the full width of the postulated tophat in [Fe/H] and age.

Given the tightness of the RGB, for the first three populations we
adopt a full-width for the [Fe/H] spread of 0.1 dex, while for the
first through third we find acceptable fits by adopting age spreads of
2.8, 2.4, and 2.0 Gyr, respectively.  The SGB of the
third population (the presence of which is clearly required by the
upper main-sequence stub above the SGB of the second population at
{\mv} $\sim~2.7$) is not obvious, resulting from the diffusing effect
of its age spread on the stars in the CMD.

For the fourth population we adopt spreads of 0.1 dex in [Fe/H] and
1.2 Gyr in age, which do not well-produce the observations, leading as
noted above to a need for alternative phenomena to improve the fit.

\item 

One aspect of our four populations that appears to be at strong odds
with the results of high-resolution spectroscopic analysis of Carina
red giants is the existence of stars having abundances in the range
--2.9 $<$  [Fe/H] $\leq$ --2.5 (\citealp{koch08a} (2), \citealp{venn12}
(2), \citealp{lemasle12} (1)) (5 unique stars; some 10\%
of the sample), which have metallicities considerably below that of
our first (oldest and most metal-poor) population which has [Fe/H] =
--1.8.

In our discussion in Section~\ref{sec:tightness} of available [Fe/H]
abundances for Carina red giants, we argued that the
Ca~II-infrared-triplet based [Fe/H] abundances of \citet {koch06} and
\citet{starkenburg10} use calibrations which are not applicable to
Carina, and suggested it would be better to derive [Fe/H] abundances
based on model atmosphere line strength calculations of the triplet,
together with consistent model atmosphere parameters (in particular,
ages) derived from isochrones.  These criticisms notwithstanding, we
note for completeness that \citet{starkenburg10} report that some 10\%
of Carina's RGB stars have [Fe/H] $<$ --2.5. 

We shall return to this discussion in Paper II (Section 10). 

\item 

The four populations have different radial distributions.  The
concentrations of the first three increase as one goes from the
(oldest) first population to the (more recent) third one. The
concentration of the (youngest) fourth population is, however, less
than that of any of the older ones. The simplest explanation is that
there is little if any direct physical connection between the fourth
population and the other three.  We conjecture that the metallicity of
the fourth population is significantly lower than our adopted value
of [Fe/H] $\sim$ --1.15.
 
\item 

The mass fractions of the four populations based on stars currently
within an elliptical radius of 13.1{$\arcmin$} and extrapolated to
their times of formation by assuming a Salpter mass function, were (in
decreasing age order) 0.34, 0.39, 0.23, and 0.04.

\item 

In Paper II we shall present chemical abundances for a sample of 63
Carina RGB stars based on model atmosphere analysis of high-resolution
spectroscopy.  Comparison in Section~\ref{sec:tightness} of the
position of these stars in the CMD with those of radial-velocity
selected samples of \citet{koch06} and \citet{walker09} shows that the
spectroscopically selected sample is incomplete at the 5 -- 10\%
level, in the sense that it does not include a complete sample of
hotter upper giant branch stars.  We noted that these bluer objects
are most likely AGB stars.  We also conjectured in
Section~\ref{sec:raddis} that they may, in some part, be first-RGB
members of our fourth (and youngest) population.  

\item

Based on the parameters of the four populations of our synthetic CMDs,
we present predicted distribution functions for [Fe/H], {\alphafe},
and age for stars on Carina's upper RGB which provide insight into the
relative contributions of the four populations.  We shall return to
these in Paper II, where we compare them with the results of our
investigation of the chemical abundances of our 63 star sample of
Carina's RGB.

\end{itemize}

\acknowledgements

We are very pleased to thank P.B. Stetson for making the photometry of
Carina available, and to acknowledge very helpful discussions with
G.S. Da Costa and A.D. Mackey.  Studies at RSAA, ANU, of the Galaxy's
dwarf satellite systems by J.E.N. and D.Y. are supported by Australian
Research Council DP0663562, DP0984924, DP120100475, DP150100862, and
FT140100554.  K.A.V. acknowledges support from the Canadian NSERC
Discovery Grants program.  This work was partly supported by the
European Union FP7 program through ERC grant number 320360.

\newpage


\begin{figure}[htbp!]
\vspace{+3.0cm}
\hspace{+1.7cm}
\includegraphics[width=10.10cm,angle=-90]{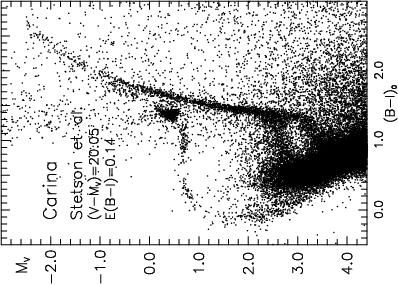}
\hspace{+20.0cm}

\vspace{-10.10cm}
\hspace{+9.7cm}
\includegraphics[width=6.5cm,angle=0]{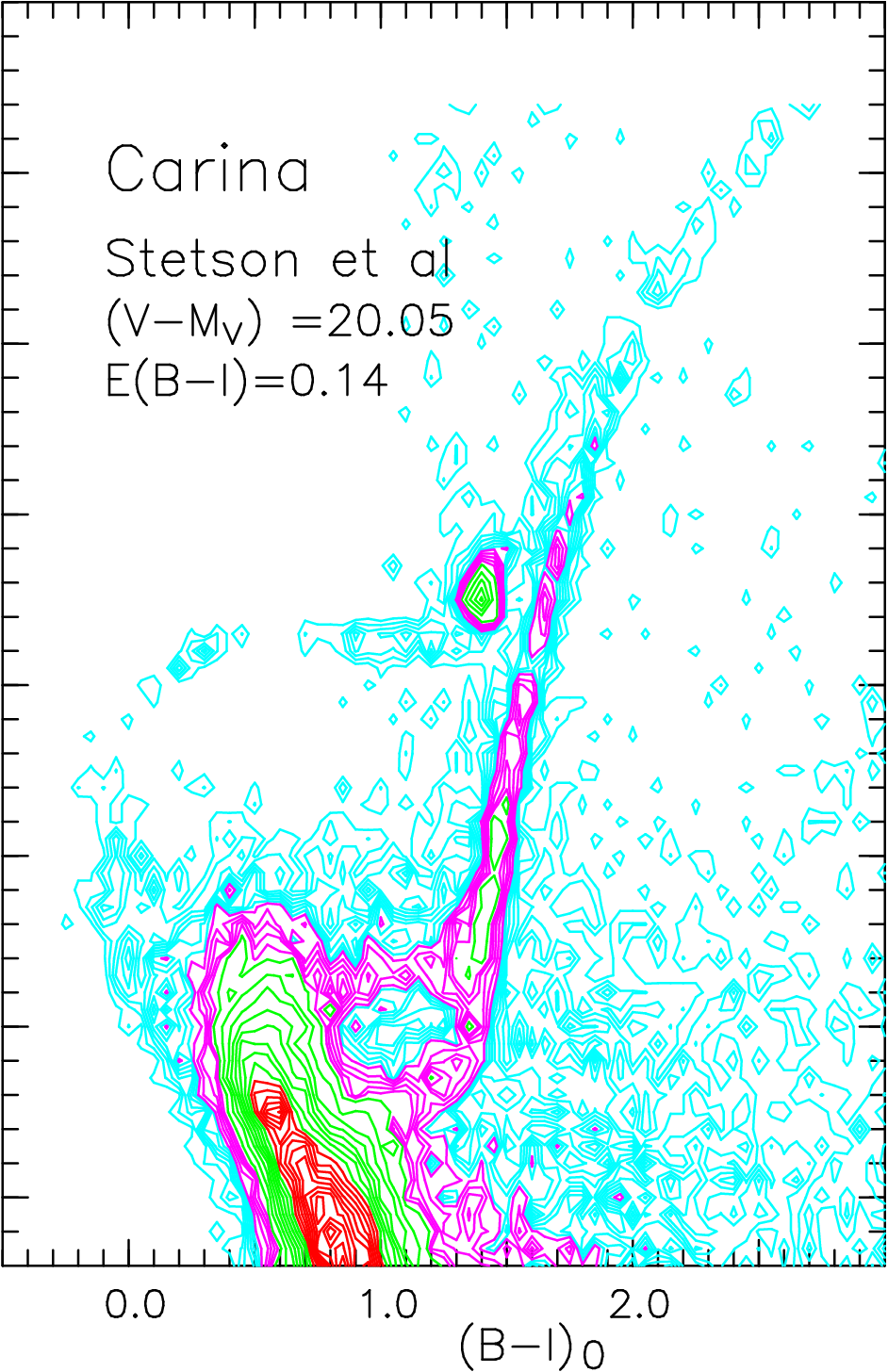}
  
\caption{\label{fig:carcmd}\small \small The color-magnitude and Hess
  diagrams of Carina, based on data provided by P.B. Stetson (2014,
  private communication).  The distance modulus and reddening follow
  \citet{venn12}.   }

\end{figure}


\begin{figure}[htbp!]
\begin{center}

\includegraphics[width=7.0cm,angle=-90]{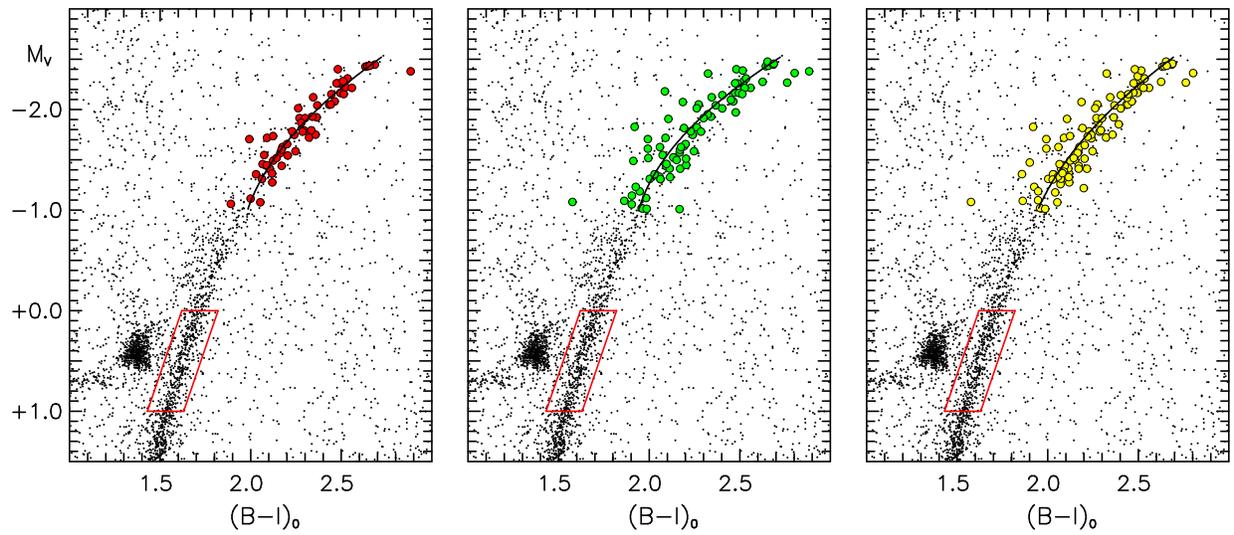}

  \caption{\label{fig:uppercmd}\small The CMD of the upper RGB of
    Carina, where all symbols present data from Stetson (2014, private
    communication). On the left the large red circles represent the
    radial-velocity members of Paper II; in the middle the
    large green symbols stand for those of \citet{koch06}; and on the
    right the large yellow symbols are based on the values of
    \citet{walker09}.  The red boxes define a region for which the
    width of the giant branch has been determined.  See text for
    discussion.}

\end{center}
\end{figure}


\begin{figure}[htbp!]
\begin{center}

\includegraphics[width=14.0cm,angle=0]{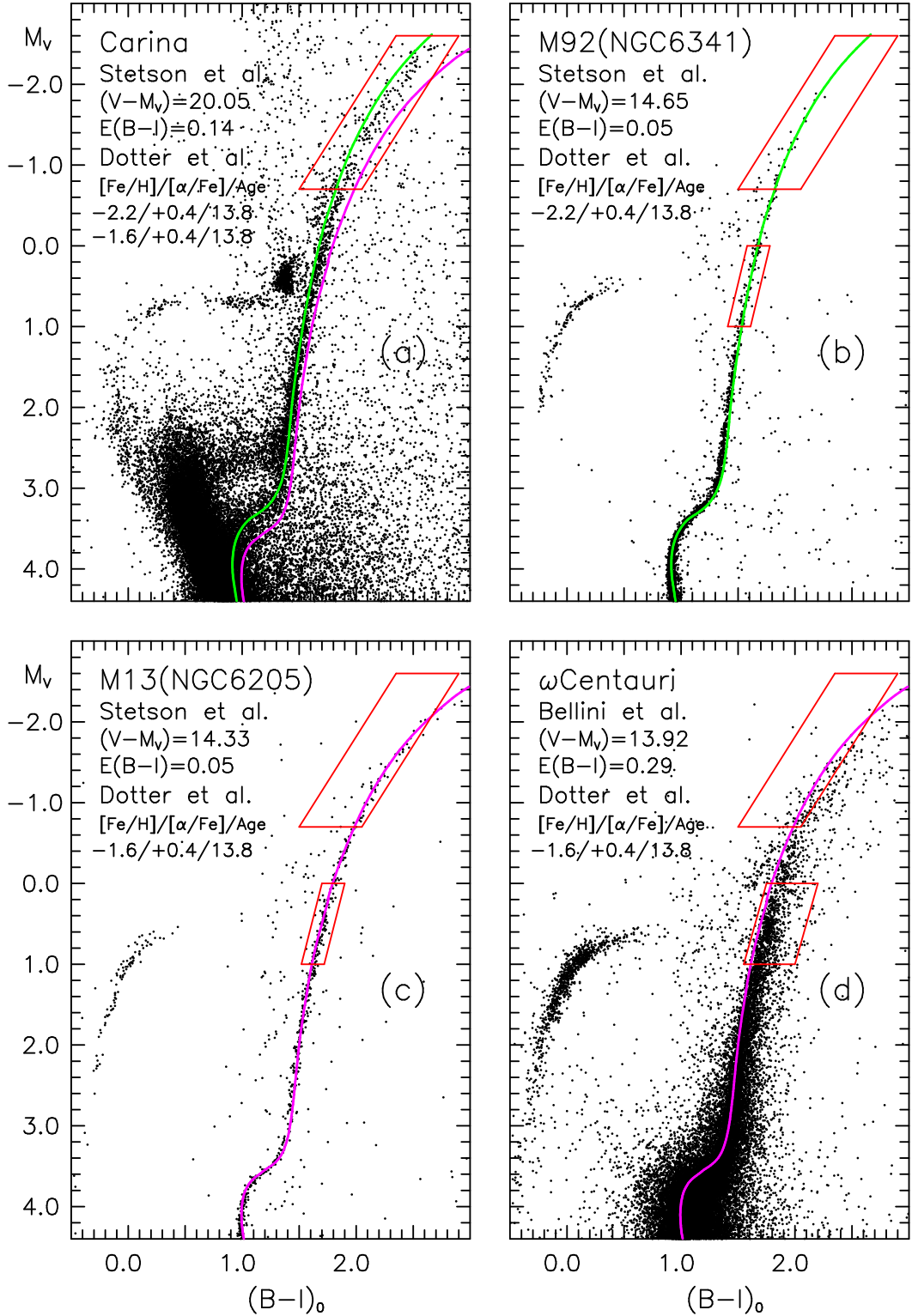}

  \caption{\label{fig:cargc}\small Comparison of the color-magnitude
    diagram of Carina (panel a) with those of M13 (panel c), M92
    (panel b), and $\omega$~Centauri (panel d). For M13 and M92 the
    data have been taken from Stetson
    (http://www.cadc-ccda.hia-iha.nrc-cnrc.gc.ca/en/community/STETSON/standards/),
    while for {\wcen} they come from \citet{bellini09}.  Also
    overplotted are isochrones from \citet{dotter08} for the
    [Fe/H]/{\alphafe}/Age values included in the legends.}

\end{center}
\end{figure}

  
\begin{figure}[htbp!]
\begin{center}

\includegraphics[width=15.0cm,angle=0]{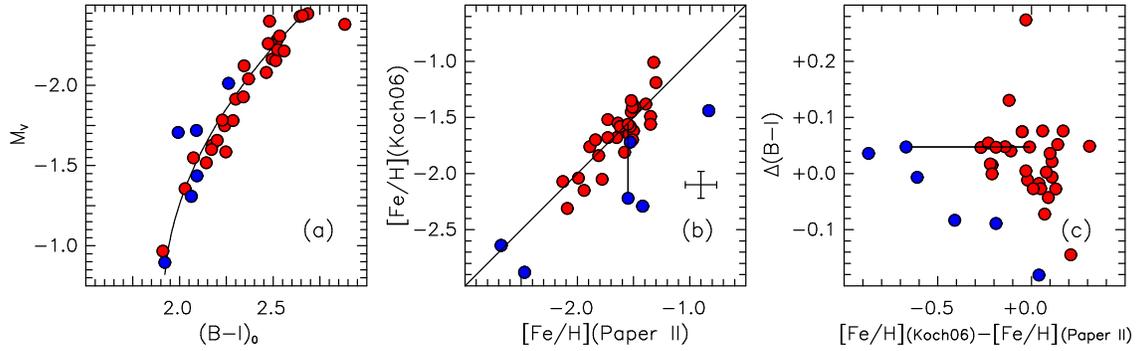}
 
 \caption{\label{fig:norkoch}\small Comparison of the [Fe/H] values of
   \citet{koch06} (determined by using the infrared Ca II triplet)
   with the [Fe/H] values of Paper II (based on high-resolution
   spectra).  Panel (a) presents the ($M_{\rm V}$, {\bio}) CMD,
   together with the quadratic line of best fit; (b) shows
   [Fe/H]$_{\rm Koch06}$ vs. [Fe/H]$_{\rm Paper II}$; and (c) plots
   $\Delta(B-I)_{0}$ (the distance a star falls from the fiducial RGB
   in the CMD in panel (a)) as a function of the abundance difference
   [Fe/H]$_{\rm Koch06}$ -- [Fe/H]$_{\rm Paper II}$ between the two
   investigations.  The two points joined by the line represent a star
   presented twice by \citet{koch06}.  See text for a discussion of
   the blue symbols.}
    
\end{center}
\end{figure}


\begin{figure}[htbp!]
\begin{center}

\includegraphics[width=14.0cm,angle=-90]{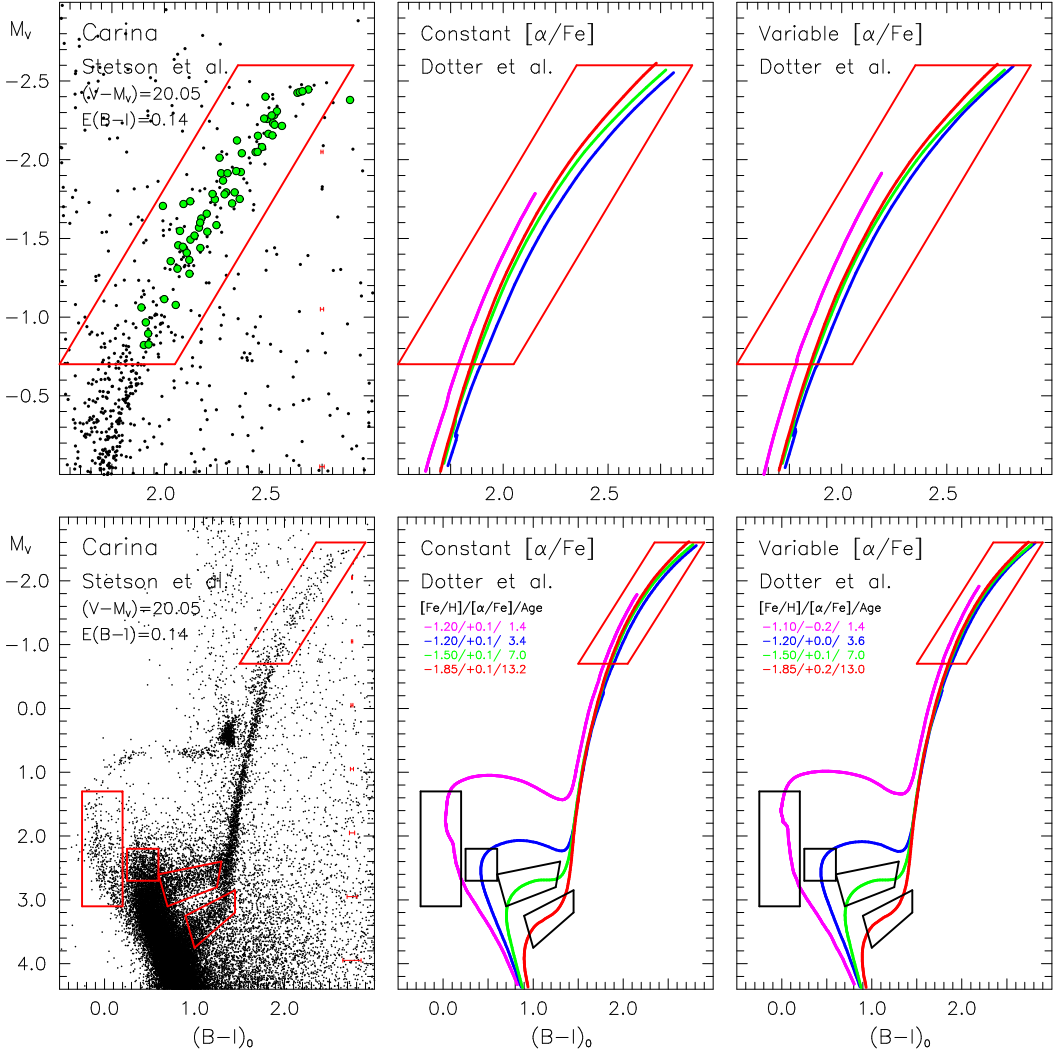}
\caption{\label{fig:cariso} \small Left panels: the observed CMD of
  Carina as shown in Figure~\ref{fig:carcmd}, superimposed in the
  lower panel with four boxes chosen to isolate unique samples of the
  four basic Carina populations.  In increasing brightness from the
  bottom of the figure these are: (i) the lower subgiant branch (SGB)
  (at {\mv} $\sim~3.3$) of the oldest population (the first
  population); (ii) the upper SGB (at {\mv} $\sim~2.7$) of the
  intermediate-age population (second population); (iii) the main
  sequence stub above the upper SGB (at {\mv}, (\bio) = (2.3, 0.5)),
  of a younger population (third population); and (iv) the upper,
  bluer, and somewhat ephemeral main sequence (at ({\mv}, \bio) =
  (2.0, 0.0) -- (3.0, 0.2)) of the youngest population (fourth
  population). The fifth and uppermost box in the figure isolates
  Carina's upper RGB sample, and (potentially) contains members of all
  of the above four putative basic populations. In the upper left
  panel the scale is expanded to permit inspection of the upper RGB in
  more detail.  In both leftmost panels, average magnitude and color
  error estimates, determined from the values tabulated by Stetson
  (2014, private communication) are included to the right.  Middle and
  right panels: isochrones from \citet{dotter08} that pass in a
  representative manner through the four population boxes.  In the
  middle panels we hold {\alphafe} constant, while in the right panels
  we allow it to vary, decreasing with decreasing age.  The population
  parameters [Fe/H], {\alphafe}, and age are shown in the legends.  }

\end{center}
\end{figure}


\begin{figure}[htbp!]
\begin{center}

\includegraphics[width=10.cm,angle=0]{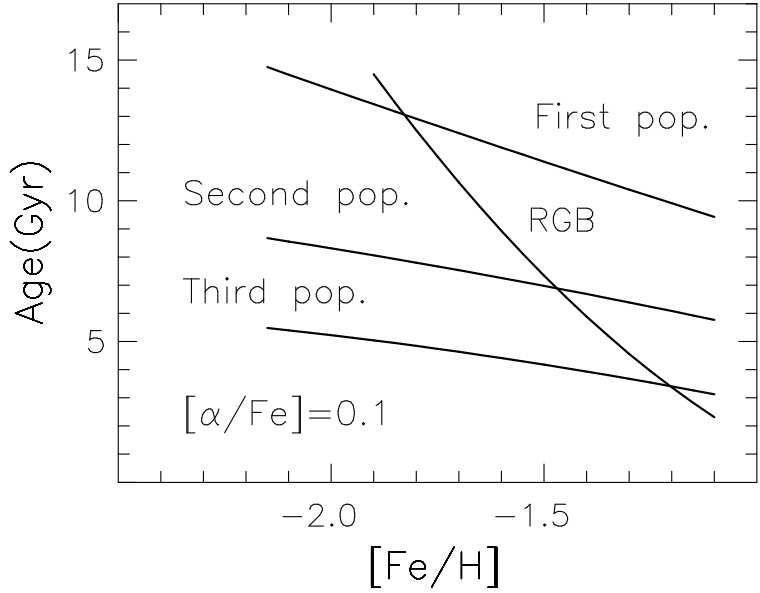}
 
 \caption{\label{fig:loci}\small Loci of the (age, [Fe/H]) values for
   isochrones having {\alphafe} = 0.1 that reproduce the four
   observable parameters (position of the SGB for the first two
   populations, the main-sequence stub of the third population, and
   the collective RGB at {\mv} = 0.5), and are consistent with the
   positions of the first three populations of the Carina CMD. The
   loci are generated by the requirement that the basic isochrones
   pass though ({\mv}, {\bio}) = (3.25, 1.20) (first population),
   (2.72, 0.97) (second population), (2.50, 0.42) (third population),
   and (0.5, 1.62) for the RGB.  The three intersection points of the
   four loci define the values pertinent to the first three epochs of
   Carina.  See text for discussion.}

\end{center}
\end{figure}


\begin{figure}[htbp!]
\begin{center}

\includegraphics[width=15.0cm,angle=-90]{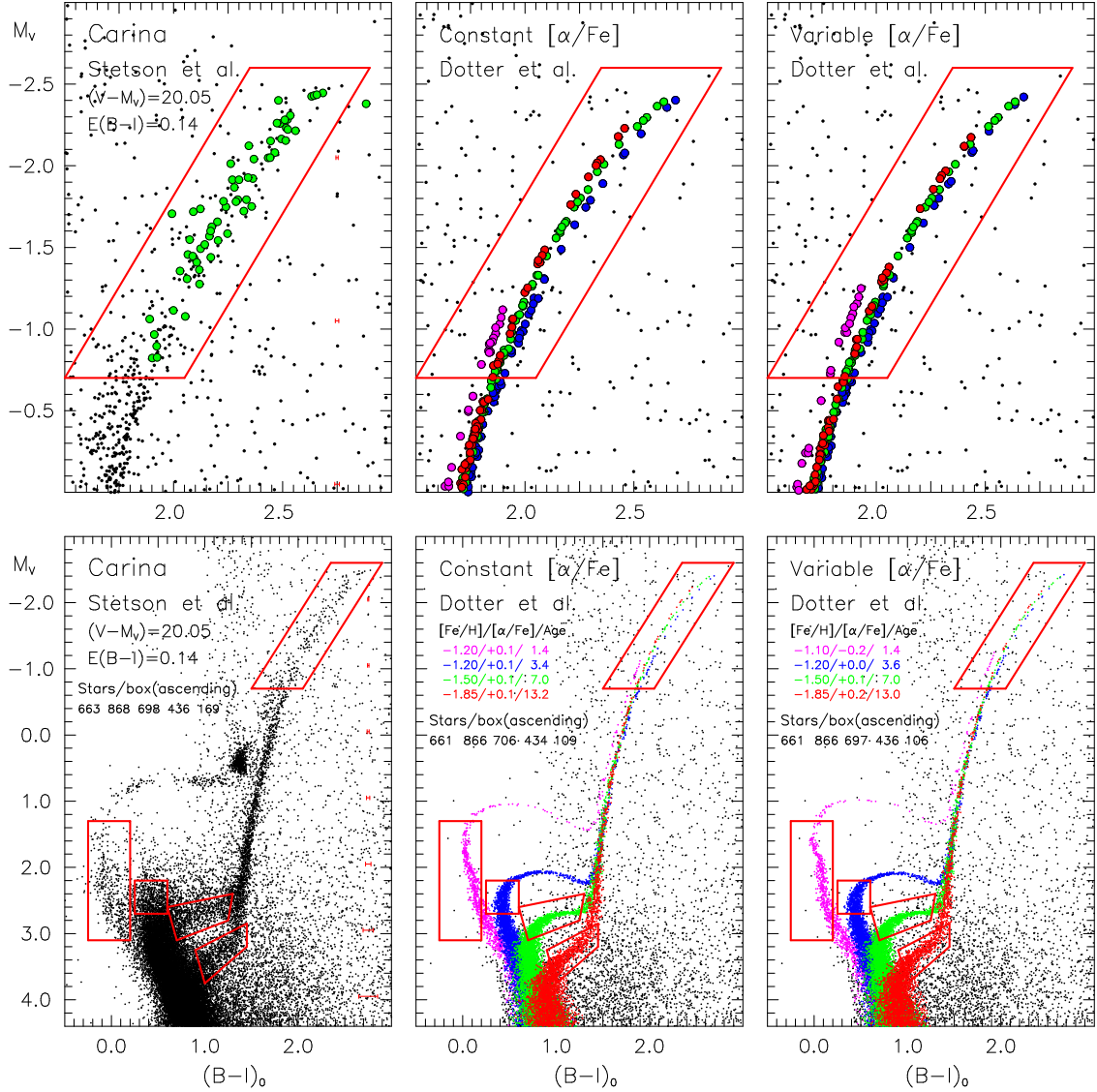}

  \caption{\label{fig:carsyn1}\small The left panels present the
    observed CMD of Carina as shown in Figure~\ref{fig:cariso},
    superimposed in the legends of the lower panel with the observed
    numbers of stars contained in each of the five boxes therein.  In
    the middle and right panels we present synthetic CMDs prepared
    using the \citet{dotter08} isochrones, and adopting magnitude and
    color errors determined from the Stetson (2014, private
    communication) tabulated uncertainties, as described in the text.
    In the middle panels, we hold {\alphafe} constant, while in the
    right we allow {\alphafe} to vary, as described in the panel
    legends.  In the legend of the lower leftmost panel are the
    numbers of stars observed in the five sub-samples, corrected for
    background interlopers by using a nearby field 25{$\arcmin$} from
    the center of Carina.  In the middle and rightmost lower panels
    the numbers refer to the model results.  Note the existence of the
    subgiant branch at {\mv} $\sim$ 2.2 in the synthetic CMDs, which
    is not seen in the observations.  See text for discussion.}

\end{center}
\end{figure}


\begin{figure}[htbp!]
\begin{center}

\includegraphics[width=15.0cm,angle=-90]{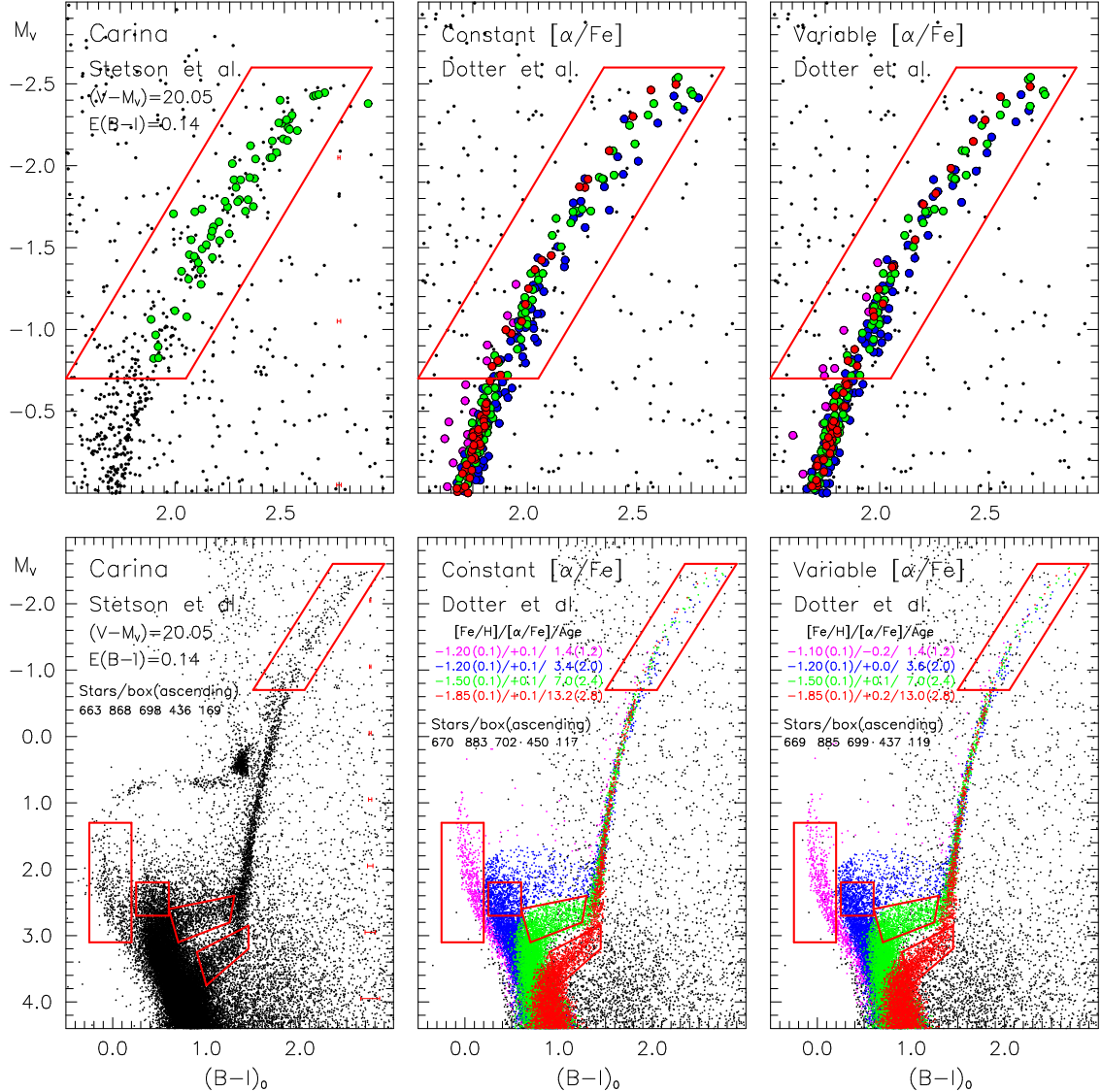}

  \caption{\label{fig:carwobble1}\small Observed and synthetic CMDs,
    with the same format as used in Figure~\ref{fig:carsyn1}.  The
    synthetic CMDs in the middle and right panels differ from those in
    Figure~\ref{fig:carsyn1} in that the synthetic magnitudes and
    colors have been modified to also randomly simulate the effect of
    flat ranges in [Fe/H] and age about the basic values, by adopting
    the small ranges shown in parentheses in the legend, as discussed
    in the text. Note that in the lower panels, these putative spreads
    in [Fe/H] and age have dispersed the subgiant branch at {\mv}
    $\sim$ 2.2 commented on in the synthetic CMDs presented in
    Figure~\ref{fig:carsyn1}, to such an extent that it is no longer
    obvious as a subgiant branch.}
   
\end{center} 
\end{figure}


\begin{figure}[htbp!]
\begin{center}

\includegraphics[width=8.0cm,angle=-90]{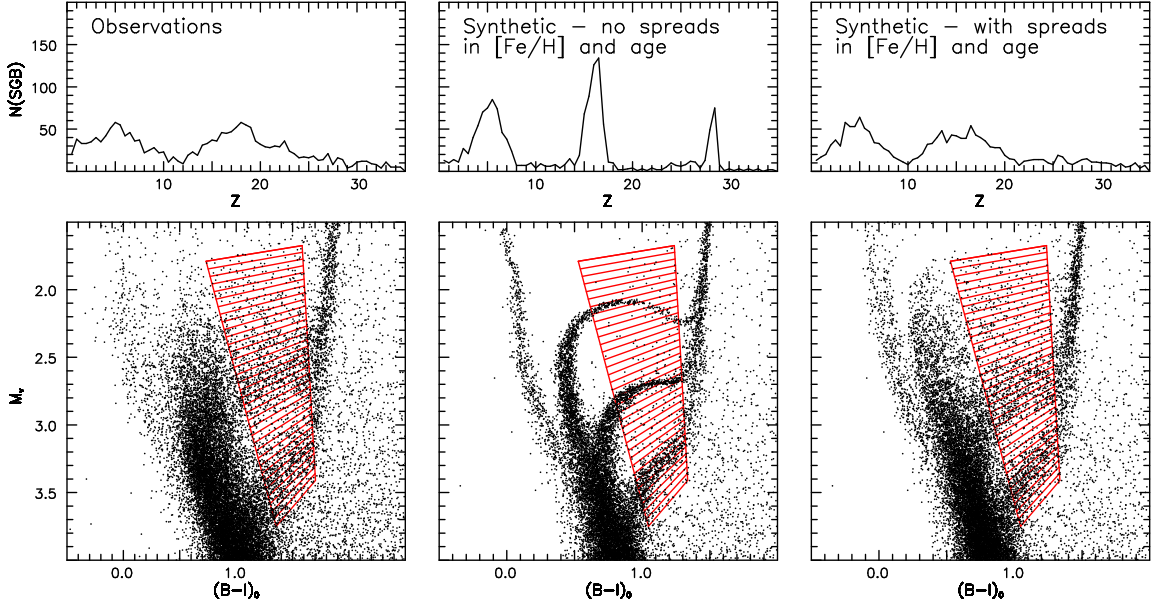}

  \caption{\label{fig:sgb}\small Lower panels: comparison between the
    observed and synthetic CMDs of Carina in the region of the SGBs of
    the first, second and third populations.  The observations are
    presented on the left, while the synthetic CMDs for the {\alphafe}
    = 0.1 case from Figures~\ref{fig:carsyn1} and ~\ref{fig:carwobble1}
    are shown in the middle (smoothing by only observational errors)
    and right (smoothing by observational errors and spreads in [Fe/H]
    and age) panels, respectively.  Upper panels: N(SRG) vs. Z, where
    N(SRG) is the number of stars in the individual elongated boxes,
    and Z represents the box number beginning at the faintest
    magnitude.}

\end{center}
\end{figure}


\begin{figure}[htbp!]
\begin{center}

\includegraphics[width=8.0cm,angle= -90]{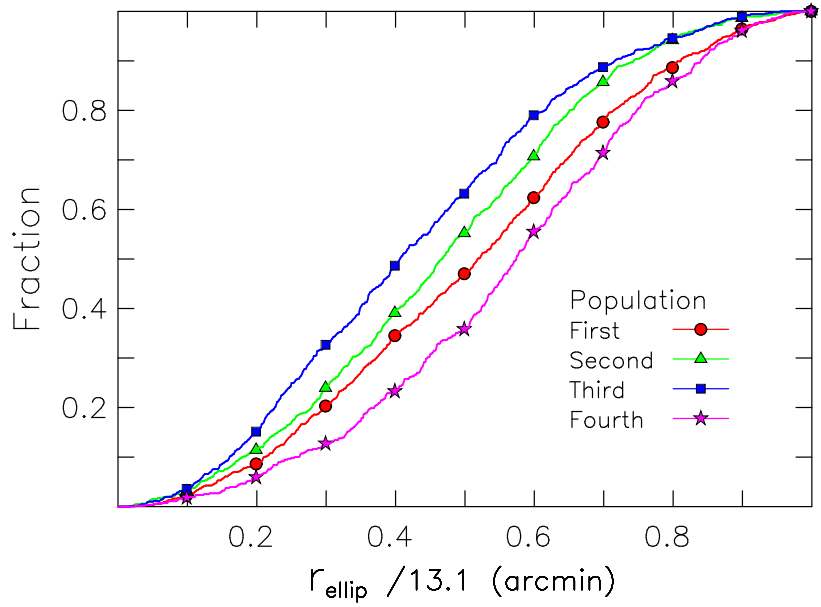}

  \caption{\label{fig:raddis}\small The cumulative distribution of the
    number of stars as a function of elliptic radius r$_{\rm ellip}$
    for Carina's four populations.  For the first three populations
    the concentrations increase as age decreases, while for the
    (youngest) fourth population, somewhat surprisingly, the
    concentration is less that for any of the other three.  See text
    for discussion.}

\end{center} 
\end{figure}


\begin{figure}[htbp!]
\begin{center}

\includegraphics[width=15.0cm,angle=-90]{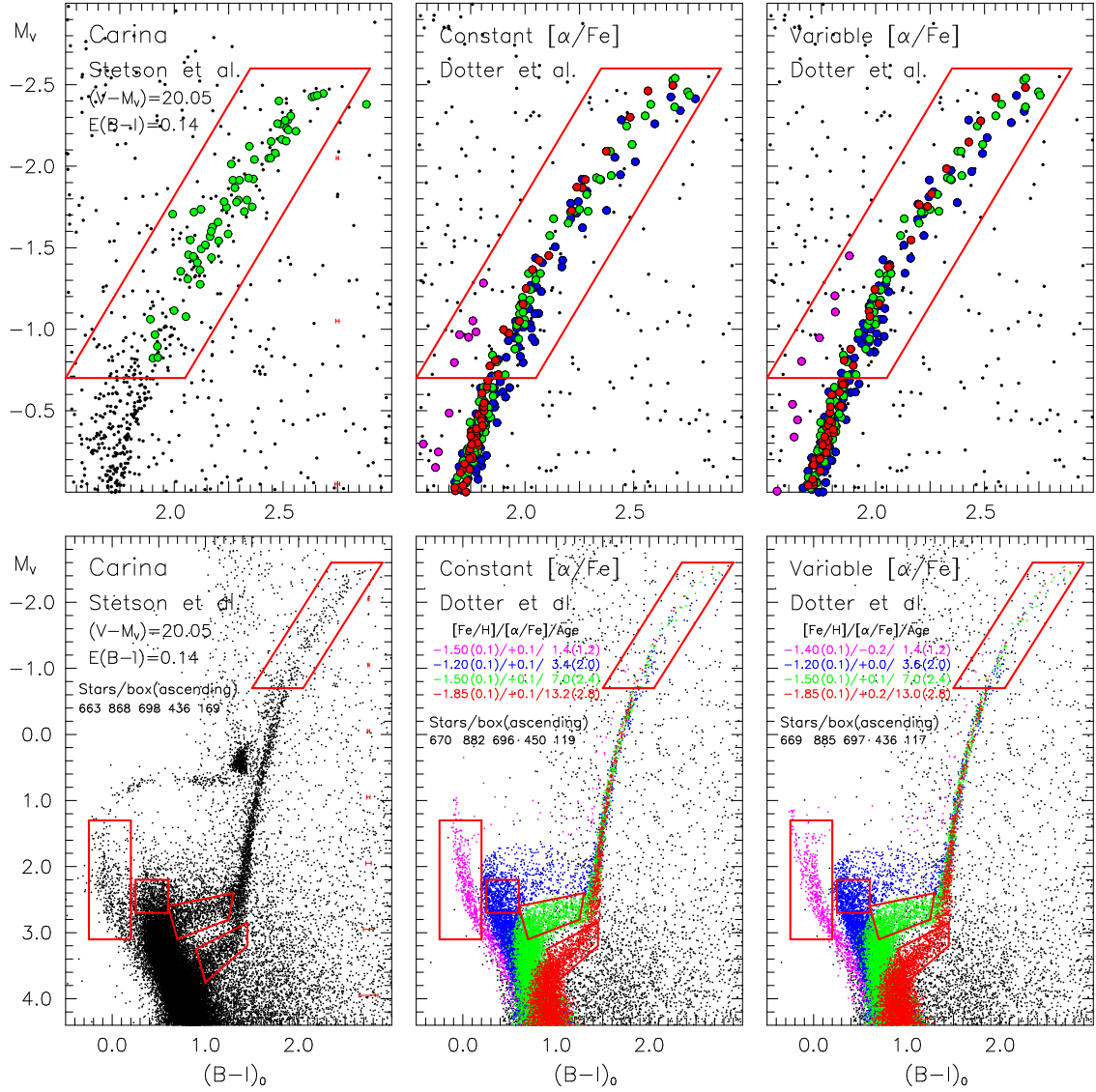}

  \caption{\label{fig:carwobble2}\small Observed and synthetic CMDs,
    with the same format as used in Figure~\ref{fig:carwobble1}.  The
    parameters of the present figure and Figure~\ref{fig:carwobble1}
    differ only in that the basic [Fe/H] values for the fourth
    population have been decreased by 0.3~dex in the former, as shown
    in the legends. There are two important consequences.  First, the
    main sequence of the fourth population is bluer by $\sim$~0.10 in
    the present figure than in Figure~\ref{fig:carwobble1}, and
    second, in the upper panels of the middle and rightmost columns of
    the present figure, there is a handful of stars of the fourth
    population bluer by {\bio} $\sim$~0.4 mag than the bulk of the RGB
    at the same {\mv}, considerably larger than that of their
    counterparts in Figure~\ref{fig:carwobble1}.  See text for
    discussion.}

\end{center}
\end{figure}


\begin{figure}[htbp!]
\begin{center}

\includegraphics[width=11.0cm,angle=0]{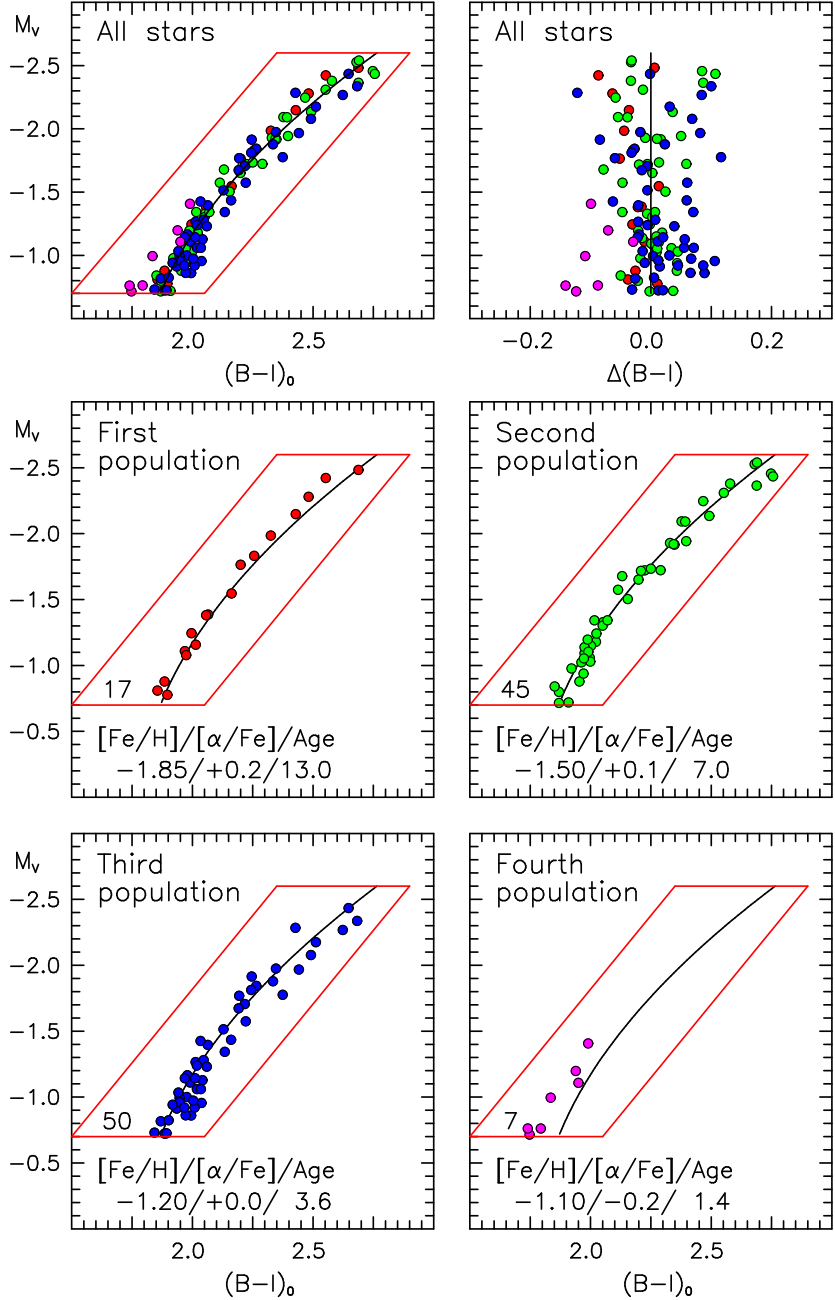}
  \caption{\label{fig:synrgb}\small The breakdown of the four
    populations on the Carina synthetic upper RGB (for the variable
    {\alphafe} case), where the red box isolates the RGB sample as
    done in previous figures, and the full lines are the quadratic
    least-squares best fits to the data.  The upper two panels present
    results for the total sample in the ({\mv}, {\bio} -- plane)
    (left) and the ({\mv}, $\Delta${\bio}) -- plane (right), where
    $\Delta${\bio}) is the distance the star fall away from the line
    of best fit in the left panel.  In the lower four panels, the
    population parameters and number of synthetic stars are presented
    for the individual populations (at bottom left of the red box.)  }

\end{center} 
\end{figure}

    
\begin{figure}[htbp!]
\begin{center}

\includegraphics[width=12.0cm,angle=-90]{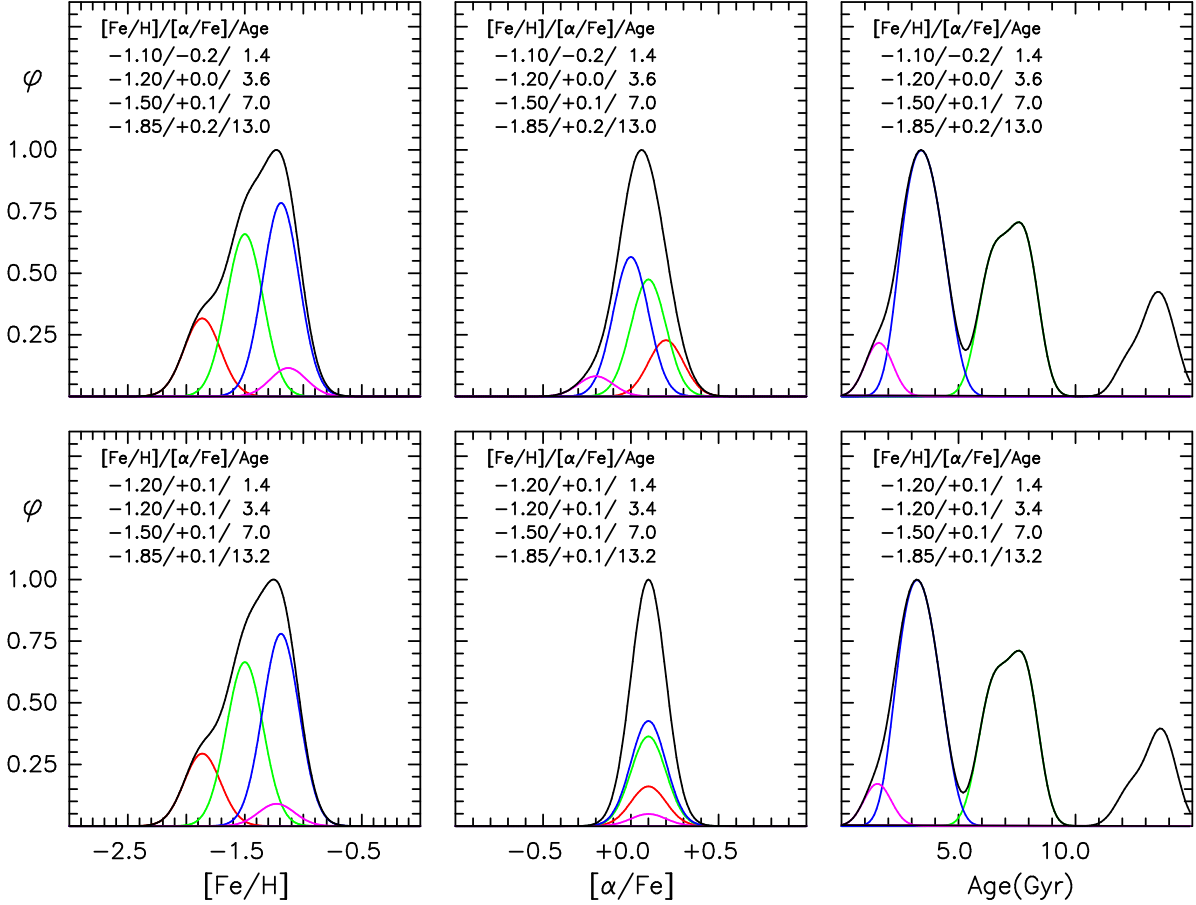}

  \caption{\label{fig:moddistrib}\small Synthetic ([Fe/H]) MDF, and
    {\alphafe} and age distribution functions of the Carina upper RGB,
    for variable {\alphafe} (upper panels) and fixed {\alphafe} (lower
    panels). Gaussian kernels having $\sigma$ values of 0.15~dex,
    0.10~dex, and 0.5~Gyr, were used for the [Fe/H], {\alphafe}, and
    age distributions, respectively.  Red, green, blue, and magneta
    refer to first through fourth populations, respectively, while
    their summation is presented in black.}

\end{center}
\end{figure}


\begin{figure}[htbp!]
\begin{center}

\includegraphics[width=5.0cm,angle=0]{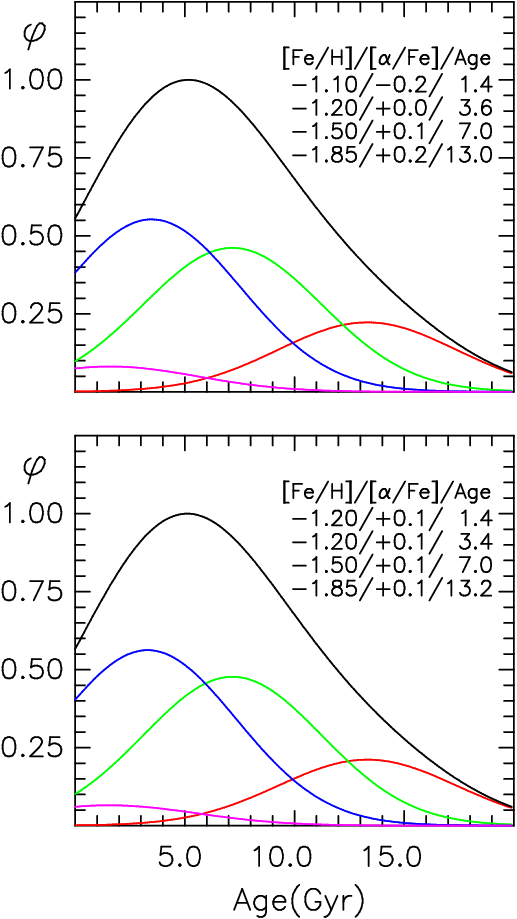}

  \caption{\label{fig:modagedist}\small Synthetic age distribution
    functions of the Carina upper giant branch for variable {\alphafe}
    (upper panels) and fixed {\alphafe} (lower panels), where a
    Gaussian kernel of width $\sigma$ = 4.0~Gyr has been adopted,
    representative of measurement errors on the RGB.  As in
    Figure~\ref{fig:moddistrib}, red, green, blue, and magneta
    refer to first through fourth populations, respectively, while
    their summation is presented in black.}

\end{center}
\end{figure}

\newpage 


\begin{deluxetable}{ll}                                                                                                                   
\tablecolumns{2}                                                                                                                            
\tablewidth{0pt}
\tabletypesize{\footnotesize}
\tablecaption{\label{tab:highlights} Major Carina Milestones}                  
\tablehead{                                                                                                                              
  \colhead  {Milestone} &                         {Authors\tablenotemark{a}} 
  } 
\startdata                                                                                                                               
The most recently discovered of the Milky Way's seven classical dSph satellite galaxies              & 1              \\ 
Carina possesses some 9 carbon stars                                                       & 2, 3, 4        \\ 
First CMD; an intermediate age galaxy                                                      & 5              \\ 
RR Lyrae variables, Anomalous Cepheids, and SX Phe stars (currently $\sim$90, 20, 430)     & 6, 7           \\  
The presence of a significant dark matter component                                         & 8, 9          \\ 
CMD accurate to the main sequence; episodic star formation; three populations              & 10, 11         \\ 
Star formation histories (2 -- 6 episodes)                                                 & 12, 13, 14, 15, 16 \\ 
Radial photometric gradients (driven by age and/or chemical abundance gradients)           & 14, 17, 18         \\ 
Large, accurate radial-velocity samples of (hundreds) of RGB members                       & 19, 20         \\ 
{\feh} estimates of large samples of RGB members based on IR Ca~II triplet                 & 19, 21         \\ 
Abundance range based on Ca~II triplet: --3. $<$ {\feh} $<$ 0., FWHM$_{\rm {\feh}} $ = 0.92 & 19             \\
High-resolution abundances for RGB samples: --3. $<$ {\feh} $<$ 0.; 0.1 $<$ {\alphafe} $<$ --0.5 & 22, 23, 24, 25, 26, 27 \\
Red giant Car-612 ``... formed in a pocket enhanced in SN Ia/II products''                 & 24             \\
Extratidal stars                                                                           & 28, 29, 30     \\
High quality CMDs                                                                          & 31, 32, 33     \\ 
Very narrow RGB sequence for a system having such large abundance and age ranges           & 31             \\ 
Star formation history (two main episodes)                                                 & 33, 34         \\ 
The Carina Project                                                                         & 7              \\ 
Comparison of CMDs with isochrones as a function of age, [Fe/H], and {\alphafe}            & 35, 36         \\ 
Chemodynamic sub-populations from medium resolution spectroscopy of $\sim$~900 stars       & 37             \\

\enddata                                                                        
\tablenotetext{a} {Sources: 1 = \citet{cannon77}; 
2 = \citet{cannon81}; 3 = \citet{mould82}; 4 = \citet{azzopardi86}; 
5 = \citet{mould83}; 
6 = \citet{saha86}; 7 = \citet{coppola15} and references therein;  
8 = \citet{mateo93}; 9 = \citet{walker09b};  
10 = \citet{smecker94}; 
11 = \citet{smecker96}; 
12 = \citet{mighell97}; 
13 = \citet{hurley98}; 
14 = \citet{hernandez00};
15 = \citet{dolphin02};
16 = \citet{monelli03};
17 = \citet{hurley01}; 18 = \citet{harbeck01};
19 = \citet{koch06};
20 = \citet{walker09};
21 = \citet{smecker99}; 
22 = \citet{shetrone03}; 
23 = \citet{koch08a}; 
24 = \citet{venn12}; 
25 = \citet{lemasle12};
26 = \citet{fabrizio12};
27 = \citet{fabrizio15};
28 = \citet{majewski05}; 29 = \citet{munoz06}; 30 = \citet{battaglia12};
31 = \citet{bono10}; 32 = \citet{stetson11}; 33 = \citet{deboer14}; 34 = \citet{santana16}; 
35 = \citet{monelli14}; 36 = \citet{vandenberg15}; 
37 = \citet{kordopatis16}
}


\end{deluxetable}      

\begin{deluxetable}{crcc}                                                         

\tablecolumns{4}        
\tablewidth{0pt}
\tabletypesize{\small}
\tablecaption{\label{tab:errors} Photometric Internal Errors}                  
\tablehead{                                    
  \colhead{$V$}   & {{\mv}}  & {$\sigma(V)$}  & {$\sigma(B-I)$} \\  
       {(1)}    & {(2)}    & {(3)}          & {(4)}            
  }  
\startdata 
  17.50   &   --2.55 &   0.001   &   0.003  \\
  18.50   &   --1.55 &   0.002   &   0.005  \\
  19.50   &   --0.55 &   0.003   &   0.008  \\
  20.50   &    0.45 &   0.004   &   0.011  \\
  21.50   &    1.45 &   0.006   &   0.017  \\
  22.00   &    1.95 &   0.008   &   0.024  \\
  22.25   &    2.20 &   0.009   &   0.029  \\
  22.50   &    2.45 &   0.010   &   0.033  \\
  22.75   &    2.70 &   0.012   &   0.040  \\
  23.00   &    2.95 &   0.016   &   0.050  \\
  23.25   &    3.20 &   0.018   &   0.059  \\
  23.50   &    3.45 &   0.021   &   0.068  \\
  23.75   &    3.70 &   0.026   &   0.079  \\
  24.00   &    3.95 &   0.031   &   0.094  \\
  24.25   &    4.20 &   0.038   &   0.111  \\
  24.50   &    4.45 &   0.042   &   0.121  \\
\enddata                                                                        

\end{deluxetable}      


\begin{deluxetable}{l|rcrc|rcr}                                                         

\tablecolumns{8}        
\tablewidth{0pt}
\tabletypesize{\small}
\tablecaption{\label{tab:pops} Basic Isochrones Used to Produce Synthetic CMDs and Deduced Mass Fractions}                  

\tablehead{                                    
\colhead  {Population}    & {[Fe/H]} & {{\alphafe}} & {Age} & {Fraction\tablenotemark{a}}  & {[Fe/H]}  & {\alphafe}   & {Age}  \\  
              {(1)}       &   {(2)}  &     {(3)}    & {(4)} & {(5)}                        &   {(6)}   & {(7)}        & {(8)} 
  }  
\startdata 
       &         & {Constant {\alphafe}}  &   &   &     &    {Variable {\alphafe}} &   \\  
First  & $-$1.85 & 0.1 & 13.2& 0.34 & $-$1.85 &  0.2   & 13.0 \\   
Second & $-$1.50 & 0.1 & 7.0 & 0.39 & $-$1.50 &  0.1   & 7.0  \\   
Third  & $-$1.20 & 0.1 & 3.4 & 0.23 & $-$1.20 &  0.0   & 3.6 \\  
Fourth & $-$1.20 & 0.1 & 1.4 & 0.04 & $-$1.10 & $-$0.2 & 1.4  \\   
\enddata                                                                        
\tablenotetext{a} {Population mass fraction (now within 13.1{$\arcmin$}) of Carina at its time of formation}   
\end{deluxetable}

\begin{deluxetable}{lcccc}                                                      

   
\tablecolumns{5}        
\tablewidth{0pt}
\tabletypesize{\small}
\tablecaption{\label{tab:litages} Multiple Population Age Estimates (Gyr) from the Literature}                  

\tablehead{                                    
\colhead  {Source}    & {First}   & {Second}   & {Third}   & {Fourth}  \\
            {(1)}     & {(2)}     & {(3)}      & {(4)   }  & {(5)}     \\    

  }  

\startdata 
\citet{smecker96}     & 11 $-$ 13  & 3 $-$ 6 &  2             &  ...  \\ 
\citet{hurley98}      & 15       & 7     &  3             &  ...  \\
\citet{hernandez00}   & 7.5      & 4.8   &  3.3           &  0.8  \\ 
\citet{dolphin02}\tablenotemark{a}    & 11.5  &  6.0   &  3.0  &  1.5 \\   
\citet{monelli03}     & 11       & 5    & {\ltsima}1     &  ...  \\
\citet{deboer14}      & $>$8     & 2 $-$ 8 &  ...           &  ...  \\
\citet{monelli14}     & 12       & 4 $-$ 8 &  ...           &  ...  \\
\citet{kordopatis16}  & 13       & 7.5     &  ...           &  ...  \\
\citet{santana16}     & $>$10    & 2 $-$ 8 &  ...           &  ...  \\
 & & & &  \\
This work             & 13       & 7       &  3.5           &  1.4    \\ 

\enddata                                                                        
\tablenotetext{a} {\citet{dolphin02} also report a fifth component with age = 0.7 Gyr} 
\end{deluxetable}      

\end{document}